\documentclass{aa}  
\usepackage{natbib}
\usepackage{graphicx}
\usepackage{footmisc}
\usepackage{float}
\usepackage{caption}
\usepackage[caption=false]{subfig}
\usepackage[colorlinks=true,citecolor=blue]{hyperref}
\usepackage{multirow}
\usepackage{amsmath}
\usepackage{nccmath}
\usepackage{color}
\usepackage{array} 
\usepackage[flushleft]{threeparttable}
\DeclareTextFontCommand{\textroman}{\fontlibertine}
\usepackage{txfonts}
\begin{document}

   \title{Na $\mathrm{I}$ and H$\mathrm{\alpha}$ absorption features in the atmosphere of MASCARA-2b/KELT-20b}

   %\subtitle{I. Overviewing the $\kappa$-mechanism}

   \author{N. Casasayas-Barris\inst{1,2} \and E. Pallé\inst{1,2} \and F. Yan\inst{3} \and G. Chen\inst{1,2,4} \and S. Albrecht\inst{5} \and L. Nortmann\inst{1,2} \and V. Van Eylen\inst{6} \and I. Snellen\inst{6} \and G.J.J. Talens\inst{6} \and J. I. González Hernández\inst{1,2} \and R. Rebolo\inst{1,2,7} \and G.P.P.L. Otten\inst{6}}
           %\fnmsep\thanks{Just to show the usage
          %of the elements in the author field}

   \institute{Instituto de Astrofísica de Canarias, Vía Láctea s/n, E-38205 La Laguna, Tenerife, Spain
              \\
              \email{nuriacb@iac.es}
         \and
             Departamento de Astrofísica, Universidad de La Laguna, Spain
        \and
            Max Planck Institute for Astronomy, Königstuhl 17, 69117 Heidelberg, Germany
         \and
             Key Laboratory of Planetary Sciences, Purple Mountain Observatory, Chinese Academy of Sciences, Nanjing 210008, China
        \and
            Stellar Astrophysics Centre, Department of Physics and Astronomy, Aarhus University, Ny Munkegade 120, DK-8000 Aarhus C, Denmark
        \and
            Leiden Observatory, Leiden University, Postbus 9513, 2300 RA, Leiden, The Netherlands
        \and
            Consejo Superior de Investigaciones Científicas, Spain}

             %\thanks{The university of heaven temporarily does not
                     %accept e-mails}

   \date{Received Month 00, 2017; accepted Month 00, 2017}

% \abstract{}{}{}{}{} 
% 5 {} token are mandatory
 
  \abstract{
  We have used the HARPS-North high resolution spectrograph ($R=115~000$) at TNG to observe one transit of the highly irradiated planet MASCARA-2b/KELT-20b. Using only one transit observation, we are able to clearly resolve the spectral features of the atomic sodium (Na I) doublet and the H$\alpha$ line in its atmosphere, which are corroborated with the transmission calculated from their respective transmission light curves.
  In particular, we resolve two spectral features centered on the atomic sodium (Na I) doublet position with an averaged absorption depth of $0.17\pm0.03\%$ for a $0.75~\mathrm{\AA}$ bandwidth, with line contrasts of $0.44\pm0.11\%$ (D2) and $0.37\pm0.08\%$ (D1). The Na I transmission light curves have been also computed, showing a large Rossiter-McLaughlin (RM) effect, with a $0.20\pm0.05\%$ Na I transit absorption for a $0.75~\mathrm{\AA}$ passband, consistent with the absorption depth value measured from the final transmission spectrum. We observe a second feature centered on the H$\alpha$ line with $0.6\pm0.1\%$ contrast, and an absorption depth of $0.59\pm0.08\%$ for a $0.75~\mathrm{\AA}$ passband, with consistent absorptions in its transmission light curves, which corresponds to an effective radius of $R_{\lambda}/R_P=1.20\pm0.04$.
  While S/N of the final transmission spectrum is not sufficient to adjust different temperature profiles to the lines, we find that higher temperatures than the equilibrium ($T_{eq}=2260\pm50~\mathrm{K}$) are needed to explain the lines contrast. Particularly, we find that the Na I lines core require a temperature of $T=4210\pm180~\mathrm{K}$ and that H$\alpha$ requires a temperature of $T=4330\pm520~\mathrm{K}$.
  MASCARA-2b, like other planets orbiting A-type stars, receives a large amount of UV energy from its host star. This energy excites the atomic hydrogen and produces H$\alpha$ absorption, leading to the expansion and abrasion of the atmosphere. The study of other Balmer lines in the transmission spectrum would allow the determination of the atmospheric temperature profile and the calculation of the lifetime of the atmosphere with escape rate measurements. In the case of MASCARA-2b, residual features are observed in the H$~\beta$ and H$~\gamma$ lines, but they are not statistically significant. More transit observations are needed to confirm our findings in Na~${\rm I}$ and H${\alpha}$, and to build up enough S/N to explore the presence of H$~\beta$ and H$~\gamma$ planetary absorptions.

  }


   \maketitle
%
%-------------------------------------------------------------------

\section{Introduction}

After the first detection of an exoplanet atmosphere by \citet{2002ApJ...568..377C}, which revealed the presence of atomic sodium (Na I) thanks to space-based instruments, several studies have been carried out using ground-based facilities which have been slowly overcoming the limitations imposed by the telluric atmospheric variations, resulting in detections of spectral signatures arising from Rayleigh scattering, Na and K with low spectral resolution  spectrographs (e.g. \citealt{2012Sing}, \citealt{2014Murgas}, \citealt{2015Wilson}, \citealt{2016Nortmann}, \citealt{2017Chena}, \citealt{2017Palle}). Observations with high resolution spectrographs are able to deal with the still remaining limitations due to the telluric atmosphere by resolving the spectral lines and taking advantage of the different Doppler velocities of the Earth, the host star and the exoplanet. 

One of the most studied species in the upper atmosphere of hot exoplanets is the Na I doublet (D2 at $\lambda5889.951~\mathrm{\AA}$ and D1 at $\lambda5895.924~\mathrm{\AA}$) due to its high cross-section. The first ground-based detection of Na I was performed by \citet{2008ApJ...673L..87R} using the High Resolution Spectrograph (HRS), with $\mathcal{R}{\sim60~000}$, mounted on the $9.2$ m Hobby-Eberly Telescope. Shortly after, \citet{2008Snellen} confirmed the Na I in HD~209458b atmosphere using the High Dispersion Spectrograph (HDS), with $\mathcal{R} = \lambda/\Delta\lambda{\sim45~000}$, on the $8$ m Subaru Telescope. Recently, ground-based studies with the High Accuracy Radial velocity Planet Searcher (HARPS) spectrograph ($\mathcal{R}{\sim115~000}$), at ESO $3.6$ m telescope in la Silla (Chile) and at $3.58$ m Telescopio Nazionale Galileo (TNG) in Roque de los Muchachos Observatory (ORM) (la Palma), have been able to resolve the individual Na I line profiles of HD~189733b \citep{2015A&A...577A..62W}, WASP-49b \citep{2017A&A...602A..36W} and WASP-69b \citep{2017CasasayasB}. Using these same data sets  \citet{2015Heng} were able to study the existence of temperature gradients, \citet{2015ApJ...814L..24L} explored the high-altitude winds of HD~189733b, \citet{2016MNRAS.462.1012B} studied stellar activity signals, and \citet{2017A&A...603A..73Y} analyzed the center-to-limb variation (CLV) effect in the transmission light curve of this exoplanet. In the near future, high-dispersion methodologies will be very important in order to observe spectral features of temperate rocky planets and super-Earths, which will be out of reach of the James Webb Space Telescope (JWST), but available with the upcoming facilities such as ESPRESSO on the Very Large Telescope (VLT) or HIRES on E-ELT (\citealt{2013Snellen}; \citealt{2017Lovis}).

Here, we present the results from only one transit observation of MASCARA-2b, obtained with the HARPS-North spectrograph (la Palma). MASCARA-2b (\citealt{MASCARA22017Talens}; \citealt{2017Lund}) is a hot Jupiter ($R_P=1.83\pm 0.07 R_J$, $M_P<3.510M_J$) transiting a rapidly-rotating ($v\sin i_{\star} = 115.9\pm3.4~\mathrm{km~s^{-1}}$) A2-type star with an orbital period of $3.474119^{+0.000005}_{-0.000006}~\mathrm{days}$ (see Table~\ref{table:planet_param} for details). This star, MASCARA-2 (also named as KELT-20 or HD~185603), is the fourth brightest star ($m_v = 7.6$) with a transiting planet known to date. With the combination of the early spectral type, the brightness and the rapid rotation of its host star, MASCARA-2b is an ideal target for transmission spectroscopy. 

\renewcommand{\thefootnote}{\fnsymbol{footnote}}
\begin{table}[h]
\small
\centering
\caption{Physical and orbital parameters of MASCARA-2b.}
\label{my-label}
\begin{tabular}{lll}
 Parameter & Symbol & Value \\ \hline \hline
 Stellar parameters \\ \hline
 \\[-1em]
 Identifiers & -&KELT-20, HD~185603 \\
 V-band magnitude &$m_{\mathrm{V}}$& $7.6$\\
 Spectral Type &-&A2\\
 Effective temperature & $T_{\mathrm{eff}}$& $8980^{+90}_{-130}~\mathrm{K}$\\
  \\[-1em]
 Projected rotation speed$^($\footnotemark[1]$^)$ & $v\sin i_{\star}$ & $115.9\pm3.4~\mathrm{km~s^{-1}}$\\
 Surface gravity & $\log g$ & $4.31\pm0.02~\mathrm{cgs}$\\
 Metallicity & [Fe/H] & $-0.02\pm0.07$\\
 Stellar mass & $M_{\star}$ & $1.89^{+0.06}_{-0.05}~\mathrm{M_{\odot}}$\\
 \\[-1em]
 Stellar radius &$R_{\star}$& $1.60\pm0.06~\mathrm{R_{\odot}}$\\ 
 \hline
 Planet parameters \\\hline
 \\[-1em]
 Planet mass$^($\footnotemark[1]$^)$ & $M_p$ &  $<3.510~\mathrm{M_{Jup}}$\\
 Planet radius & $R_p$ & $1.83\pm0.07~\mathrm{R_{Jup}}$ \\
 Equilibrium temperature &$T_{eq}$& $2260\pm50~\mathrm{K}$\\ 
 Surface gravity$^($\footnotemark[1]$^)$ &$\log g_P$& $<3.460~\mathrm{cgs}$\\ 
 \\[-1em]
 \hline
 System parameters \\\hline
 \\[-1em]
 Right Ascension& -&  19$^h$38$^m$38.73$^s$\\
 Declination & -&  +31$^\mathrm{o}$13'09.2"\\
 Epoch& $T_{c}$ & $2457909.5906^{+0.0003}_{-0.0002}~\mathrm{BJD}$\\
  \\[-1em]
 Period& $P$ & $3.474119^{+0.000005}_{-0.000006}~\mathrm{days}$\\
  \\[-1em]
 Transit duration& $T_{14}$ & $3.55\pm0.03~\mathrm{hours}$\\
 Semi-major axis &$a$& $0.057\pm0.006~\mathrm{AU}$\\
 Scaled semi-major axis &$a/R_{\star}$ &$7.5\pm0.04$\\
 Inclination & $i$ & $86.4^{+0.5}_{-0.4}~\mathrm{^o}$\\
  \\[-1em]
 Eccentricity& e & $0$ (fixed)\\
 Planet-to-star ratio & $R_p/R_{\star}$ &$0.1133\pm0.0007$\\
 Systemic velocity &$\gamma$& $-21.07\pm0.03~\mathrm{km~s^{-1}}$\\
 Projected obliquity &$\lambda$& $0.6\pm4~\mathrm{^o}$\\\lasthline
\end{tabular}
\\
\begin{tablenotes}
\item Notes. $^($\footnotemark[1]$^)$ From \citet{2017Lund}. All the remaining parameters are taken from \citet{MASCARA22017Talens}. 
\end{tablenotes}
\label{table:planet_param}
\end{table}
\renewcommand{\thefootnote}{\arabic{footnote}}

MASCARA-2b is one of the few exoplanets transiting an A-type star known to date with effective temperatures higher than ${\sim}7000~\mathrm{K}$. These planets typically receive a large amount of extreme ultraviolet radiation from its host star, exciting the atomic hydrogen to produce H$\alpha$ absorption and leading to the expansion and possibly abrasion of their atmosphere \citep{2016Burrier}. The study of the Balmer lines of these planets allows to estimate the lifetime of the atmosphere from the escape rate measurement. Comparative studies of this planet population, which present different system properties, will help us to understand their origin and evolution.

%--------------------------------------------------------------------
\section{Observations}
We observed one transit of MASCARA-2b on 16 August 2017 using the HARPS-North spectrograph mounted on the $3.58~\mathrm{m}$ Telescopio Nazionale Galileo (TNG), located at the Observatorio del Roque de los Muchachos (ORM, La Palma). For this transit, the observations started at 21:21 UT and finished at 03:56 UT, with an airmass variation from $1.0$ to $2.1$.

The observations were carried out exposing continuously before, during, and after the transit in order to retrieve a good baseline, which is key for the data reduction process. We used fiber A on the target and fiber B on the sky in order to correct possible emission features from the Earth's atmosphere. Since MASCARA-2 is a bright star, with a magnitude of $7.6$ (V), we exposed $200~\mathrm{s}$ per exposure, recording a total of $90$ consecutive spectra, $58$ of them during the transit. The individual retrieved spectra have a signal-to-noise ratio (S/N) ranging from $40$ to $80$ per wavelength bin in the continuum near the Na I doublet. One spectrum of the rapid-rotator telluric standard HR~7390 ($5.59$ V) was also observed before MASCARA-2b observations, using $300~\mathrm{s}$ of exposure time, with an airmass of $1.0$, and obtaining a S/N of ${\sim170}$ in the continuum. HR~7390 is a bright A0V-type star with a projected rotation speed of $150~\mathrm{km/s}$.

%------------------------------------------------------------- 
\section{Methods}
\label{method}
\begin{figure}[h]
\centering
\includegraphics[width=0.49\textwidth]{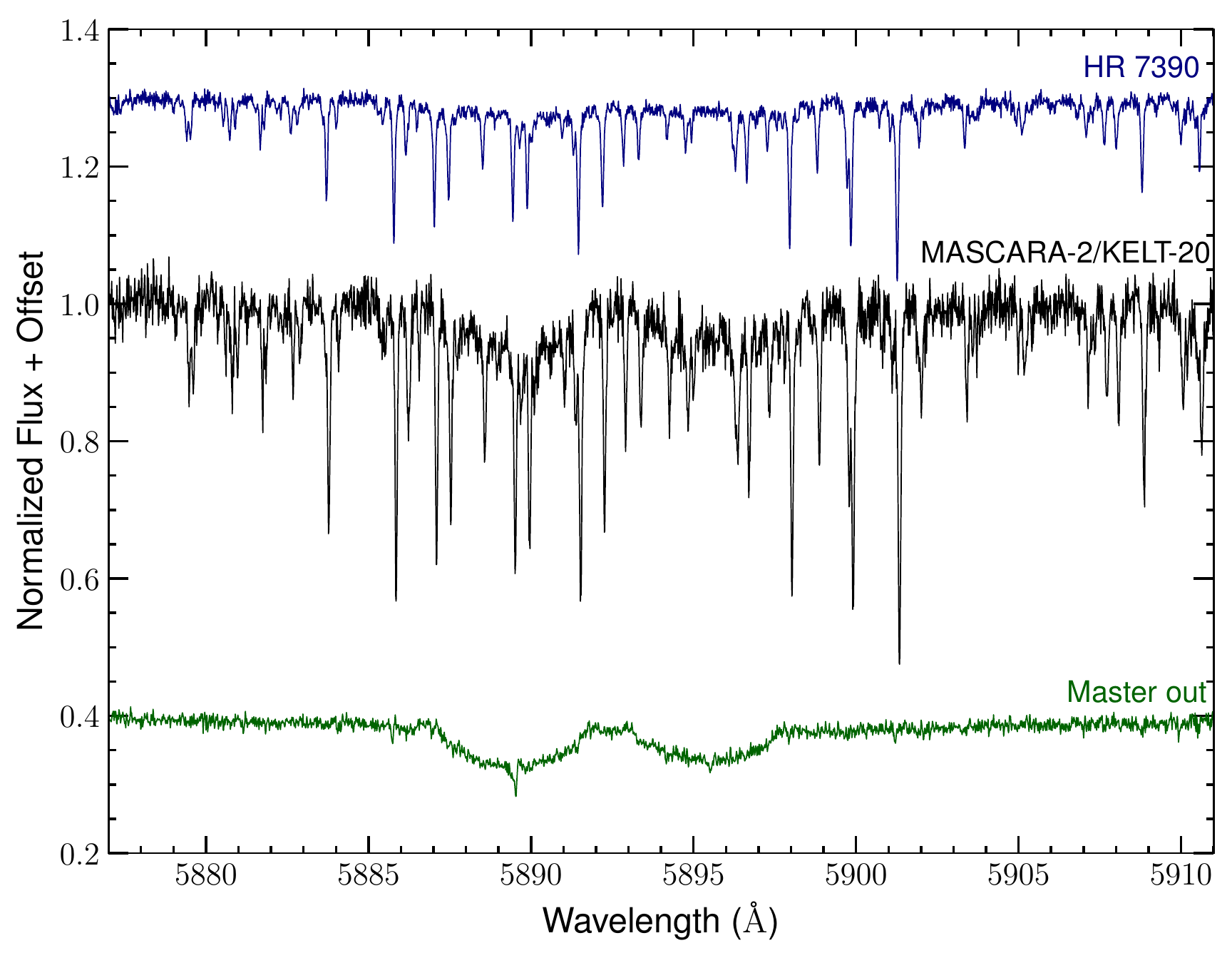}
\caption{Normalized spectra of the telluric standard HR~7390 (blue) and MASCARA-2 (black) around the Na I region after their reduction with the HARPS DRS. In green we show the MASCARA-2 master out spectrum, after the telluric correction.}
\label{fig:example_spec}
\end{figure}

The observations were reduced with the HARPS-North Data reduction Software (DRS), version 1.1. The DRS extracts the spectra order-by-order, which are then flat-fielded using the daily calibration set. A blaze correction and the wavelength calibration are applied to each spectral order and, finally, all the spectral orders from each two-dimensional echelle spectrum are combined and resampled into a one-dimensional spectrum ensuring flux conservation. The resulting one-dimensional spectra covers a wavelength range between $3800~\mathrm{\AA}$ and $6900~\mathrm{\AA}$, with a wavelength step of $0.01~\mathrm{\AA}$, referred to the Solar System barycenter rest frame and in air wavelengths. One representative spectrum of MASCARA-2 and the telluric standard HR~7390 are presented in Figure~\ref{fig:example_spec}.

The first step of our analysis is the telluric contamination subtraction. Near the Na I doublet, the main telluric contributors are water and telluric sodium \citep{2008Snellen}. Each individual target sky spectra were examined to look for the presence of telluric sodium, but no sodium emission was observed. For an A-type star as MASCARA-2, the method used in \citet{2017CasasayasB} to remove the telluric features produced by water and oxygen does not work as well as for other spectral types. Thus, here we correct for this contamination by using a combination of both \citet{2015A&A...577A..62W} and \citet{2017CasasayasB} methods, taking advantage of the observations of the telluric standard HR~7390. With the method described in \citet{2013A&A...557A..56A}, which makes the assumption that the variation of the telluric lines follows linearly the airmass variation, we compute a high-quality telluric spectrum. The difference here is that, before computing the high-quality telluric spectrum, we align all the telluric lines by using the Barycentric Earth Radial Velocity (BERV) information from the HARPS-N file headers. Then, instead of scaling all the MASCARA-2 spectra to the averaged in-transit airmass as in \citet{2015A&A...577A..62W}, we scale them to the telluric standard airmass (i.e. $1.0$ in this case). Finally, and after removing the small stellar features from the telluric standard spectrum as in \citet{2000stellarremove}, each scaled MASCARA-2 spectrum is divided by the final telluric standard spectrum, which has been also shifted to the Earth's reference frame.

Once the telluric contamination has been removed, we follow the steps described in \citet{2017CasasayasB} to extract the transmission spectrum of the planet. We note that the stellar radial velocity (RV) information of MASCARA-2, as for other rapidly-rotating A-type stars, is not well determined by the HARPS-N pipeline. For rapidly-rotating early-type stars, which present very broad lines in their spectrum, the stellar RV correction is not very important and the possible uncorrected RV shift between the spectra do not result in a large difference, for this reason, we do not consider the stellar RV correction here. However, if the stellar lines are deep and narrow, the RV correction becomes important. In this way, the possible instrumental RV effects are not considered, but such effects produce offsets which similarly affect all spectra, resulting in a global wavelength shift with respect the expected rest frame position.

The combination of all the out-of-transit spectra (master out) after subtracting the telluric contamination is shown in Figure~\ref{fig:example_spec}. Sharp residuals can be observed at $\sim{5890~\mathrm{\AA}}$ and at $\sim{5896}~\mathrm{\AA}$ (less deeper), which appear in each in-transit and out-of-transit spectra. The measured RV of both signals is $-14.9\pm0.5~\mathrm{km/s}$, with no wavelength shift during the observation. In order to determine their origin, we used the LISM Kinematic Calculator \citep{2008ISMtool}, which gives us information about the interstellar medium clouds that our line-of-sight traverses while observing a target. In the case of MASCARA-2b, according to the interstellar medium distribution in our Galaxy, the sight line went through the G, Mic and Oph clouds which have  $-14.10\pm0.97~\mathrm{km/s}$, $-20.90\pm1.34~\mathrm{km/s}$, $-28.31\pm0.93~\mathrm{km/s}$ RVs, respectively. The very similar RV of the G cloud and the residuals observed in the data is the evidence of their possible interstellar origin. Since the interstellar sodium does not show any shift during the night and, in this particular case, we are not correcting for the stellar RV, the presence of interstellar species is not a problem, as it is compensated when dividing the in-transit spectra by the master out spectrum.

During the observation, the planet is moving along the orbit, changing its RV with respect to the observer. This RV change results in a wavelength shift of the measured planetary absorption lines that needs to be corrected. When the in-transit spectra are divided by the master out spectrum, each of these spectra contain residuals that need to be shifted to the planetary rest frame in order to become aligned before co-adding all of them. Here, this wavelength shift is obtained by calculating the theoretical RV of the planet along the orbit, and projecting it to the observer's line of sight. MASCARA-2b is a recently discovered planet and only an upper-limit of its mass (${\sim}3.5~\mathrm{M_J}$; \citealt{2017Lund}) is known. For this reason, in order to calculate the RV we consider here that $\mathrm{M_P<<M_{\star}}$ (i.e. assuming the planetary mass does not significantly contribute to the reduced mass of the system). The calculated RV changes during the transit from $-24~\mathrm{km/s}$ to $+24~\mathrm{km/s}$, approximately. The maximum difference of this RV between considering the mass upper-limit and the $\mathrm{M_P<<M_{\star}}$ approximation is $\sim \pm 0.1~\mathrm{km/s}$, which is not significant considering that a pixel in HARPS-N corresponds to $\sim 0.8~\mathrm{km/s}$.

\begin{figure}[h]
\centering
\includegraphics[width=0.49\textwidth]{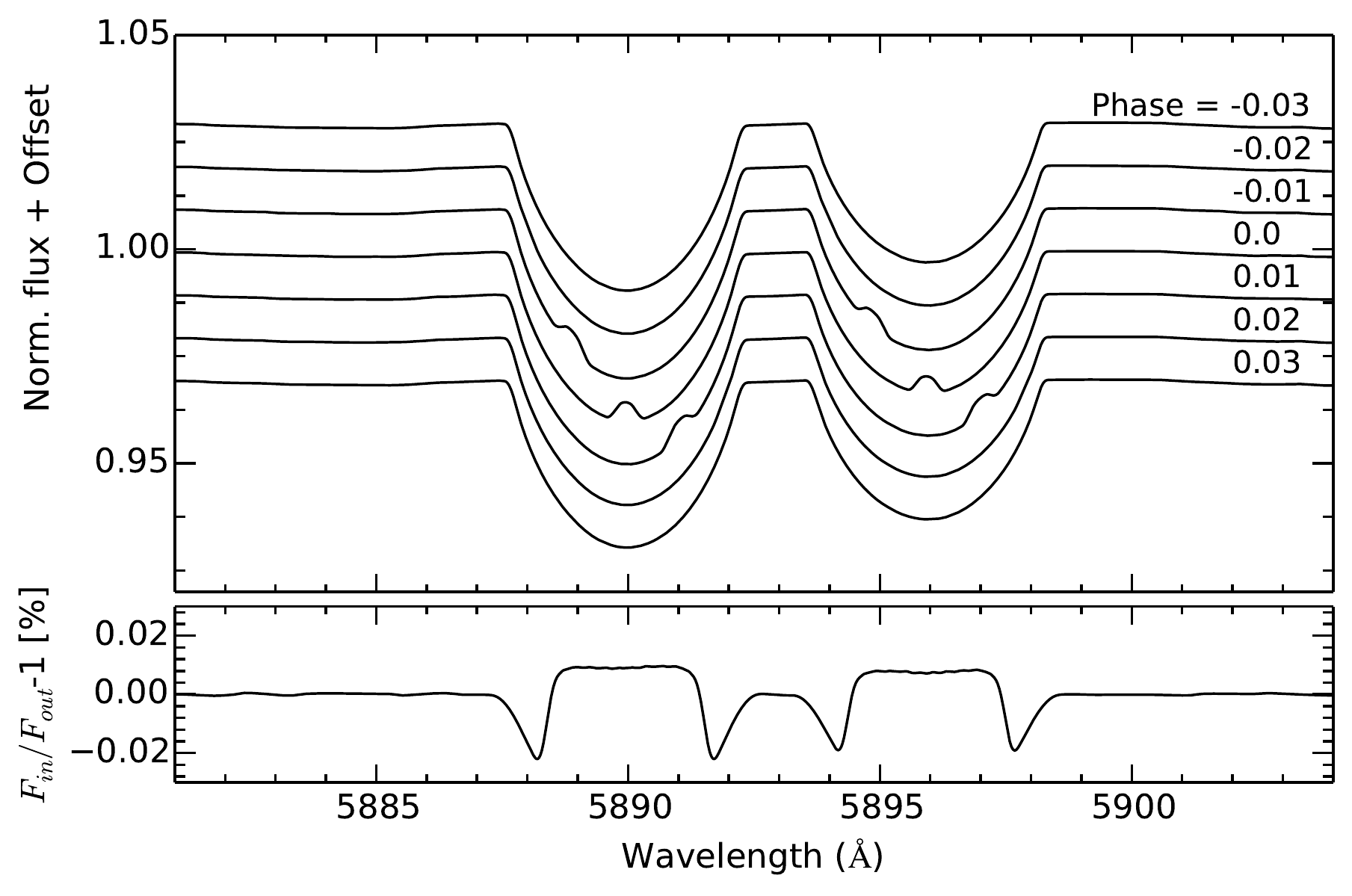}
\caption{Modeled spectra of MASCARA-2 for different orbital phases of the planet using LTE and a spin-orbit angle of $0^o$ \citep{MASCARA22017Talens} for the RM effect (top), and modeled RM and CLV effects in the final transmission spectrum (bottom). We note the large RM effect in the stellar lines.}
\label{fig:RV_lines}
\end{figure}

Finally, we model the stellar spectrum for different orbital phases of the planet using ATLAS9 models \citep{ATLAS9}. The modeled spectra contain both the center-to-limb variation (CLV) and Rossiter-McLaughlin (RM) effects on the stellar lines shape (see top panel of Fig.~\ref{fig:RV_lines}). The CLV effect is obtained by computing the stellar spectra with different $\mu$ angles as in \citet{Yanmodel} and the RM effect is obtained assuming rigid rotation with a spin-orbit angle of $0^o$ (\citealt{MASCARA22017Talens} measured $\lambda = 0.6\pm 4~\mathrm{^o}$), stellar rotation of $v\sin i = 115~\mathrm{km/s}$ and an impact parameter of $0.5$ \citep{2017Lund}. By applying the same methodology to these synthetic spectra, we obtain how the RM and CLV affect the final transmission spectrum of the planet (see bottom panel of Fig.~\ref{fig:RV_lines}). As can be observed, although the expected RM effect is large, the combination of both effects is of the order of $\sim 10^{-4}$ (i.e. one magnitude below the typical noise). This is because the CLV is very weak for these spectral-type stars and the RM is smeared when the different in-transit spectra are combined. Nevertheless, we divide the final transmission spectrum by the model to correct for these effects.

A sort of 2D representation of the original spectra and the results after the different processing steps described in this section, are presented in the Appendix~\ref{appendix_2D}. In these figures, the RM effect on the lines and the exoplanet absorption in the stellar and planetary rest frame can be seen.

%----------------------------------------------------------------- 
\section{Results}

\subsection{Transmission spectrum analysis of Na I}
\label{sec:TSA}
The S/N per extracted pixel in the continuum near the Na I doublet for the observed spectra of MASCARA-2, retrieved with HARPS-N, ranges from $40$ to $80$. When combining the spectra to obtain the master spectrum for in-transit observations and out-of-transit observations the S/N increases to $\sim{150}$, and when co-adding all the spectra to compute the final transmission spectrum it increases to $\sim{450}$.

\begin{figure*}[h]
\centering
\includegraphics[width=0.85\textwidth]{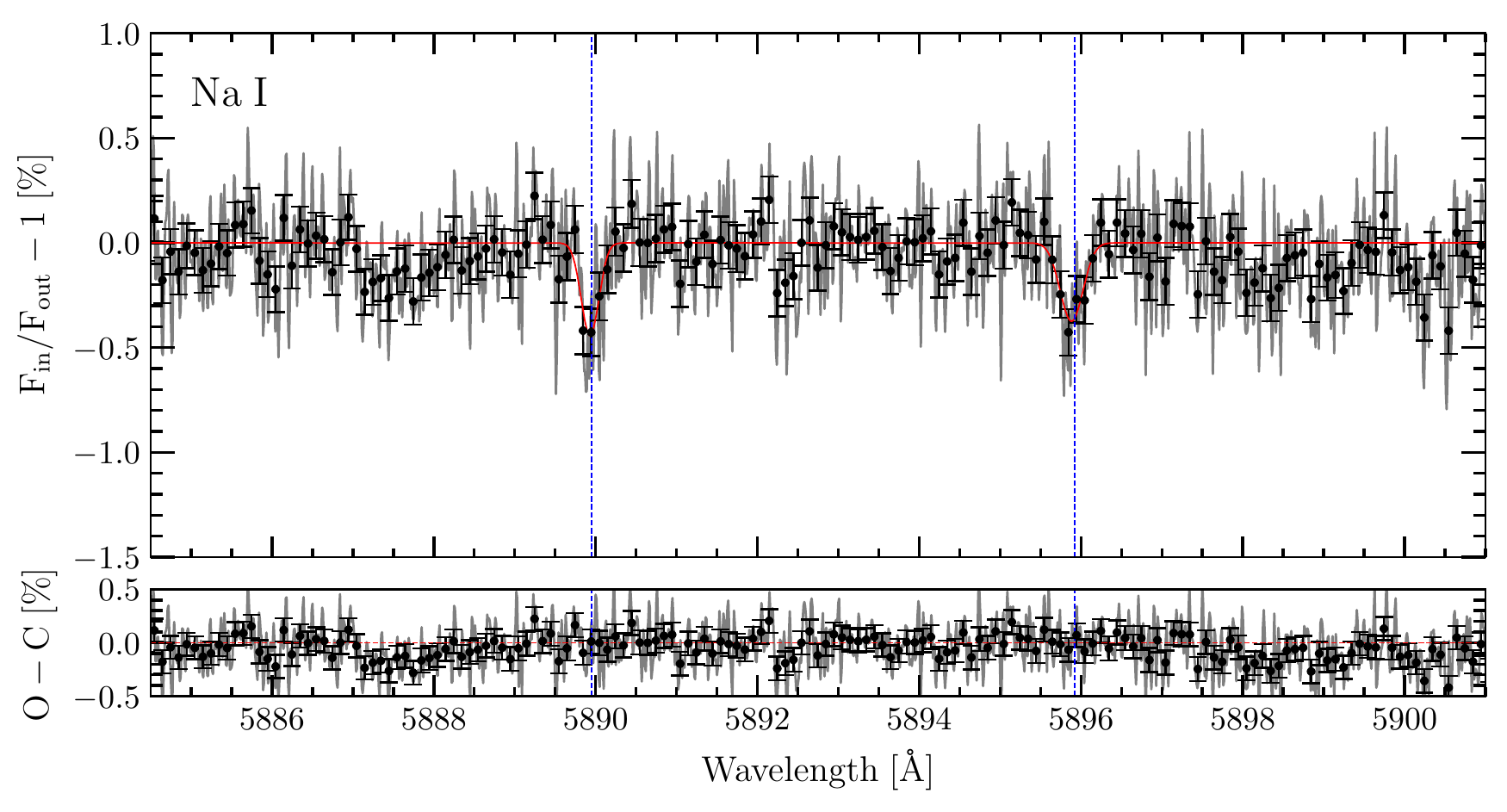}
\caption{Transmission spectrum of MASCARA-2b atmosphere in the region of Na I D doublet. In the top panel the atmospheric transmission spectrum is presented in light grey. With black dots we show the binned transmission spectrum by 10 pixels and the Gaussian fit to each Na I D lines is shown in red, with its residuals in the bottom panel. The expected wavelength position of the Na I doublet lines, in the planetary reference frame, are indicated with blue vertical lines. The uncertainties of the relative flux come from the error propagation of the photon and readout noise from the original data.}
\label{fig:TS_Na}
\end{figure*}

The final transmission spectrum of MASCARA-2b around the Na I is shown in Figure~\ref{fig:TS_Na}. As it can be observed in this Figure, both Na I D lines show a peak relative to the continuum. With a Gaussian fit to each Na I line, we measure line contrasts of $0.44\pm0.11\%$ (D2) and $0.37\pm0.08\%$ (D1), and a full width at half maximum (FWHM) of $0.26\pm0.08~\mathrm{\AA}$ (D2) and $0.33\pm0.08~\mathrm{\AA}$ (D1), centered on $5889.93\pm0.03~\mathrm{\AA}$ (D2) and $5895.88\pm0.03~\mathrm{\AA}$ (D1). The expected values for the D2 and D1 lines in the planetary rest frames are $5889.95~\mathrm{\AA}$ and $5895.92~\mathrm{\AA}$, respectively, implying that no net blue- or red-shift is measured. In both cases the Gaussian fit has a reduced $\chi ^2 $ value of ${\sim1.2}$. This fit procedure is performed using a simple Markov chain Monte Carlo (MCMC), and the best-fit values are obtained at $50\%$ (median) and their error bars correspond to the $1\sigma$ statistical errors at the corresponding percentiles. 

The absorption depth of the Na I lines is measured by calculating the weighted mean of the flux in a central passband of different bandwidths ($0.188~\mathrm{\AA}$, $0.375~\mathrm{\AA}$, $0.75~\mathrm{\AA}$, $1.5~\mathrm{\AA}$, $3.0~\mathrm{\AA}$ and $5.0~\mathrm{\AA}$) centered on each Na I lines, and then compared with the average of two $12~\mathrm{\AA}$ bandwidths taken in the continuum of the transmission spectrum of the Na I lines, one in the blue (B$~= [5872.89-5884.89]~\mathrm{\AA}$) and one in the red (R$~= [5900.89-5912.89]~\mathrm{\AA}$) regions. The final values are presented in Table~\ref{tab:AD_valuesNa}. We note that all the uncertainties are calculated with the error propagation of the photon noise level and the readout noise from the original data.

\renewcommand{\thefootnote}{\fnsymbol{footnote}}
 \begin{table*}[t]
\centering
\caption{Summary of the measured relative absorption depth in [\%] in the Na I D lines of the final transmission spectrum and the transmission light curves (TLC) of MASCARA-2b for different bandwidths.}
\begin{tabular}{lcccc}
\hline\hline
& & Na $\mathrm{I}$& & \\ 
\cline{2-5}
\\[-1em]
Bandwidth &$\mathrm{D_2}$ & $\mathrm{D_1}$& D doublet$^($\footnotemark[1]$^)$ &TLC$^($\footnotemark[3]$^)$ \\ \hline
\\[-1em]
$0.188~\mathrm{\AA}$ & $0.435\pm0.085$ &$0.386\pm0.083$&$0.411\pm0.059$ &$-$\\ \hline
\\[-1em]
$0.375~\mathrm{\AA}$&$0.320\pm0.058$ &$0.337\pm0.057$ & $0.324\pm0.041$ & $-$\\ \hline
\\[-1em]
$0.75~\mathrm{\AA}$ & $0.168\pm0.042$ & $0.178\pm0.041$ & $0.172\pm0.029$ &$0.200\pm0.046$ \\ \hline
\\[-1em]
$1.5~\mathrm{\AA}$& $0.066\pm0.030$ &$0.076\pm0.030$ &$0.071\pm0.021$ & $0.116\pm0.033$\\ \hline
\\[-1em]
$3.0~\mathrm{\AA}$ &  $0.057\pm0.022$ &  $0.045\pm0.021$ & $0.051\pm0.015$ &$0.077\pm0.024$ \\ \hline
\\[-1em]
$5.0~\mathrm{\AA}$&  $0.072\pm0.017$&$0.066\pm0.017$&$0.069\pm0.012$& $0.083\pm0.019$\\ 
\hline\hline
\end{tabular}\\
\begin{tablenotes}
\item Notes.$^($\footnotemark[1]$^)$ Average of both Na I D1 and D2 lines absorption depth. $^($\footnotemark[3]$^)$ Absorption measured in the final transmission light curves (see third row of Fig.~\ref{fig:TLC_NaI}).
\end{tablenotes}
\label{tab:AD_valuesNa}
\end{table*}
\renewcommand{\thefootnote}{\arabic{footnote}}

Different control distributions (not shown here) were performed to ensure that the results are not caused by stellar or telluric residuals, but can only be obtained with the correct selection of the in- and out-of-transit samples. In particular, we computed the transmission spectrum by only considering the in-transit spectra, being randomly selected to form the synthetic in- and out-of-transit samples. The same test was performed by considering only the out-of-transit files, and also selecting the even files as the in-transit sample and the odd ones as the out-of-transit sample. In all tests, the resulting synthetic transmission spectra are mainly flat, with clear difference with the lines depth of the real transmission spectrum. The averaged absorption depth in the expected Na~$\mathrm{I}$ D2 and D1 lines position for a $0.75~\mathrm{\AA}$ bandwidth are $0.021\pm0.055\%$, $0.058\pm0.040\%$, and $0.087\pm0.032\%$ for the "out-out", "in-in" and "even-odd" samples, respectively. The final transmission spectrum with a non-correction of the planet RV was also computed, with no signals in the Na~$\mathrm{I}$ position ($0.045\pm0.029\%$ of absorption for a $0.75~\mathrm{\AA}$ bandwidth), supporting that the possible interstellar sodium do not affect the final result and the Na $\mathrm{I}$ signals observed in the final transmission spectrum only appear in the planetary rest frame. In addition, this is the evidence that we do not observe stellar activity signals, which would be visible in the stellar rest frame if they were present \citep{2017A&A...602A..36W}. As can be observed, the "in-in" and "even-odd" distributions absorption depths, the final transmission spectra are not totally flat as in the "out-out" sample. We note that in these distributions we do not correct for the CLV+RM effects and, as can be observed in the transmission light curves, the in-transit absorption varies during the transit. These effects could introduce some variations in the absorption and not allowing the total compensation of the planet absorption in the “in-in” and “even-odd” control distributions. This does not happen in the "out-out" sample, as expected. On the other hand, the small absorption measured when the planet RV is not considered, could be produced because the residuals partially overlap in the stellar rest frame.

\begin{figure}[h]
\centering
\includegraphics[width=0.49\textwidth]{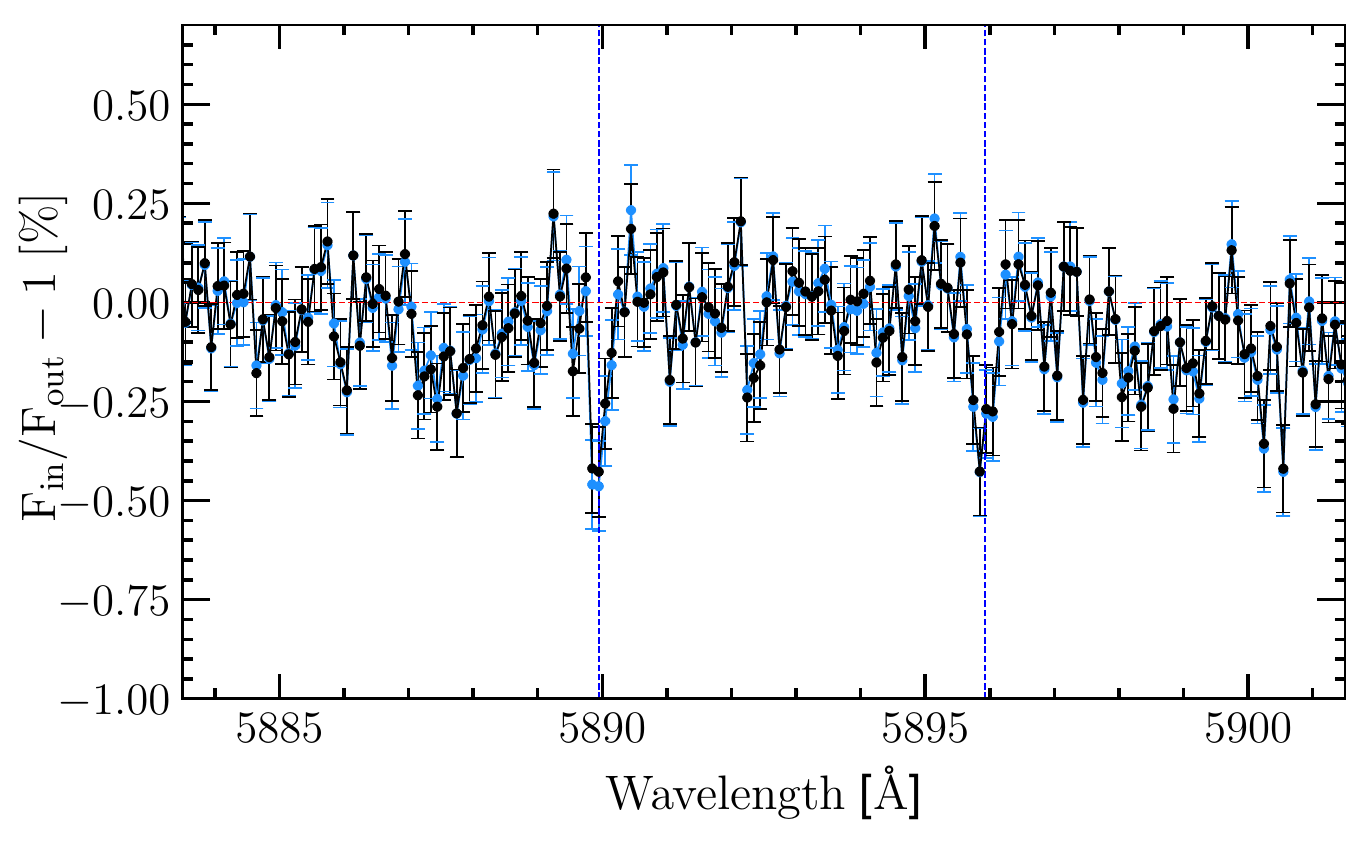}
\caption{Comparison between the transmission spectra obtained with (light blue line) and without (black line) stellar RV correction, before correcting for the final RM and CLV effects. In order to calculate the result with the stellar RV correction, we adopt the upper limit stellar RV semi-amplitude ($322.51~\mathrm{m/s}$, \citealt{2017Lund}) and a theoretical Rossiter-McLaughlin (RM) effect model calculated using \citet{2005Ohta} with the parameters presented in \citet{MASCARA22017Talens}. In both cases the spectrum is binned by 10 pixels. The blue vertical lines show the expected wavelength position of the Na $\mathrm{I}$ D lines and the red horizontal line is the null-absorption reference.}
\label{fig:TSRVnoRV}
\end{figure}

As commented in Section~\ref{method}, since no precise RV parameters have been measured for MASCARA-2b system, we tested the result with stellar RV corrected using the upper limit value of the RV semi-amplitude ($322.51~\mathrm{m/s}$, \citealt{2017Lund}) and a theoretical Rossiter-McLaughlin (RM) effect model. The resulting transmission spectrum is very similar to the result without stellar RV correction (see Fig.~\ref{fig:TSRVnoRV}). As no significant differences are observed in the measurements when the stellar RV is corrected and when it is not, we show our analysis based on the case in which the stellar RV is not contemplated as this is the most correct procedure in this particular case.

\subsection{Transmission light curve of Na I}
\label{TLCNa}
\begin{figure*}[h]
\centering
\includegraphics[width=1.\textwidth]{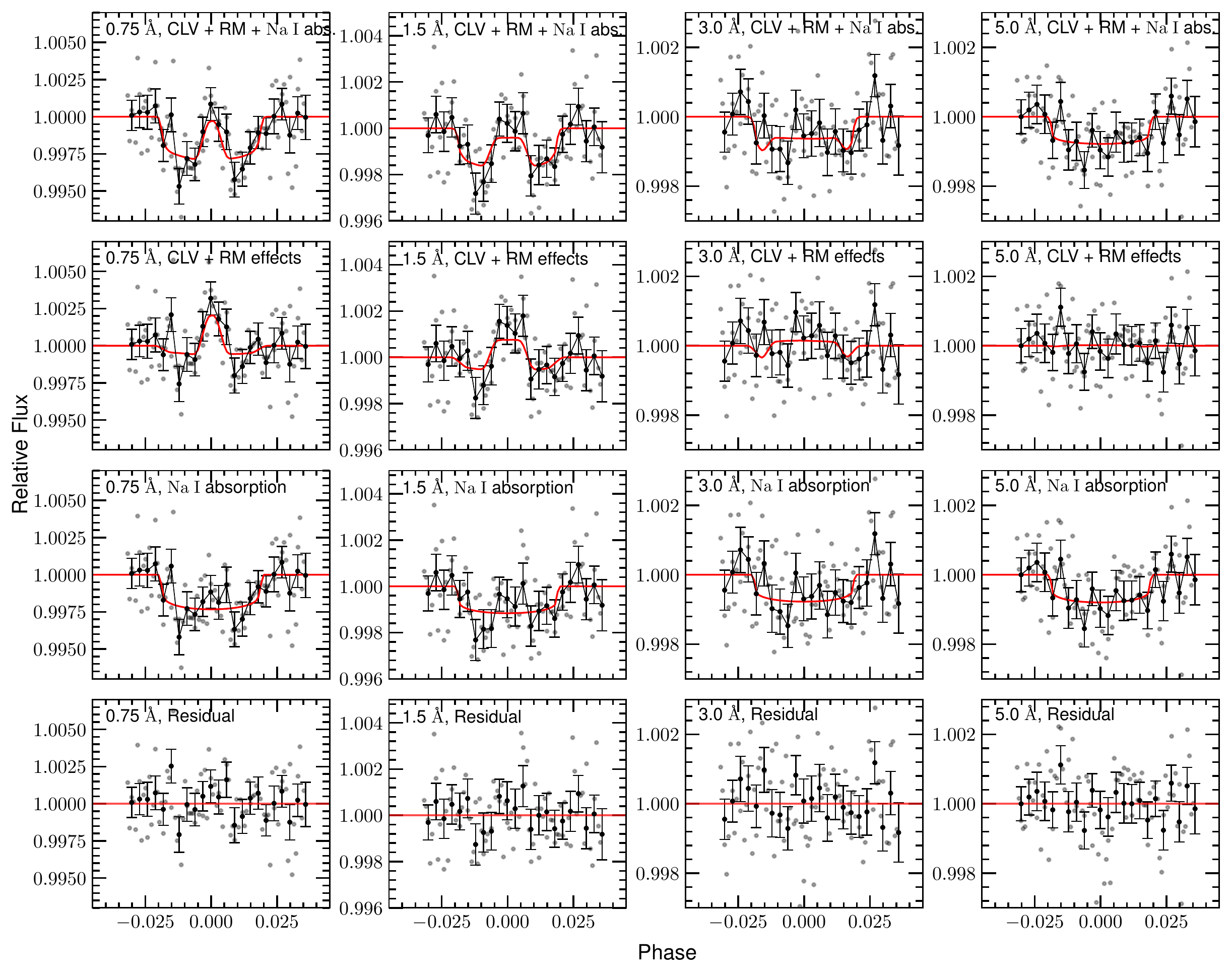}
\caption{Na I transmission light curves from HARPS-N observations of MASCARA-2b, for four different bandwidths: $0.75~\mathrm{\AA}$ (first column), $1.5~\mathrm{\AA}$ (second column), $3.0~\mathrm{\AA}$ (third column), $5.0~\mathrm{\AA}$ (fourth column). In all panels, the light gray dots are the values from all the spectra, and the black dots the same data binned with 0.003 phase step. First row: Observed transmission light curve of the Na I D lines from our data reduction. The relative flux of the D1 and D2 lines is averaged. The red line is the modeled transmission light curve, which is the combination of the best-fit Na I absorption and the CLV and RM effects \citep{2017A&A...603A..73Y}. This model is calculated theoretically and there is no fitting to the data. Second row: The CLV+RM effects. The black line is the result of removing the Na I absorption model from the observed transmission light curve. The modeled CLV and RM effects are shown in red. Third row: The Na I absorption light curve obtained dividing the observed transmission light curve by the modeled CLV+RM effects (black line). The red line is the best-fit Na I absorption model. Forth row: Residuals between the observed transmission light curve and the model. We note the different y-scale of the panels.}
\label{fig:TLC_NaI}
\end{figure*}

The transit light curve of the Na I D lines is calculated as presented in \citet{2017A&A...603A..73Y} and \citet{2017CasasayasB} using four different bandwidths: $0.75~\mathrm{\AA}$, $1.5~\mathrm{\AA}$, $3.0~\mathrm{\AA}$, $5.0~\mathrm{\AA}$. Since the stellar lines are broad because of the rapid-rotation of the star ($v\sin i_{\star} = 115~\mathrm{km/s}$) and we expect a large RM effect, we included the calculation for a broad $5.0~\mathrm{\AA}$ passband, which encompasses the full Na I line. As reference passbands we use the same wavelength regions presented in Section~\ref{sec:TSA}. The results of both D1 and D2 lines are averaged and, finally, the data are binned with a $0.003$ phase step ($\sim3.3~\mathrm{km/s}$ of planetary radial-velocity), taking into account the S/N of the data. The final observed light curves for the different bandwidths are shown in the first row of Figure~\ref{fig:TLC_NaI}.

The resulting light curves are the combination of the planetary absorption, the CLV and the RM effects. For MASCARA-2 we expect a large RM (see Fig.~\ref{fig:RV_lines} and \citealt{2017Lund}) and a small CLV effect. We note that both effects produce similar shapes in the transmission light curves \citep[see][]{2017A&A...603A..73Y}, being in this case the RM the most significant contribution. 
It is noticeable how the RM effect in the transmission light curves diminishes as the width of the central passband is increased. This is because the deformation in the stellar sodium lines due to the RM effect occurs during different phases of the transit and, for small bandwidths, it can move in and out of the reference passbands. However, the $5~\mathrm{\AA}$ passband encompasses the complete stellar sodium lines, averaging out the effect in the transmission light curves.

To correct for these effects we divide the observed transmission light curves by the light curve with the modeled effects  (see second and third row of Fig.~\ref{fig:TLC_NaI}). We note that the modeled CLV + RM light curves are directly derived from the spectral models shown in Figure~\ref{fig:RV_lines}, and the fitting procedure is only applied to the Na I absorption model. This fitting is performed by using a simple MCMC algorithm and the model built with PyTransit \citep{Parviainen2015Pytransit}. This model depends on the orbital period ($P$), the excentricity ($e$), the planet-star ratio ($R_P/R_{\star}$), the scaled semi-major axis ($a/R_{\star}$), the orbital inclination ($i$), the transit center ($T_0$), the argument of periastron ($\omega$) and the quadratic limb darkening coefficients ($u$). The $u$ coefficients are fixed to the estimated values calculated with the PyLDTk \citep{Parviainen2015LDTK} Python Package, which use the library of PHOENIX stellar atmospheres \citep{Husser2013},  $R_p/R_{\star}$ remains free and the other parameters are fixed to the values presented in Table~\ref{table:planet_param}.

As it can be observed, the CLV + RM model is slightly underestimated compared to the measurements, which is possibly due to the assumption of LTE during the model computation \citep{2017A&A...603A..73Y}. With the final Na I transmission light curve it is possible to calculate the true absorption depth by comparing the weighted mean of the in-transit and out-of-transit values (see the results in Table~\ref{tab:AD_valuesNa}). These values are consistent with the measured absorption depth values derived from the final transmission spectrum of MASCARA-2b, presented in the same table for a better comparison.

\subsection{Analysis of H$\alpha$ region}
\label{Ha}
\begin{figure*}[h]
\centering
\includegraphics[width=0.85\textwidth]{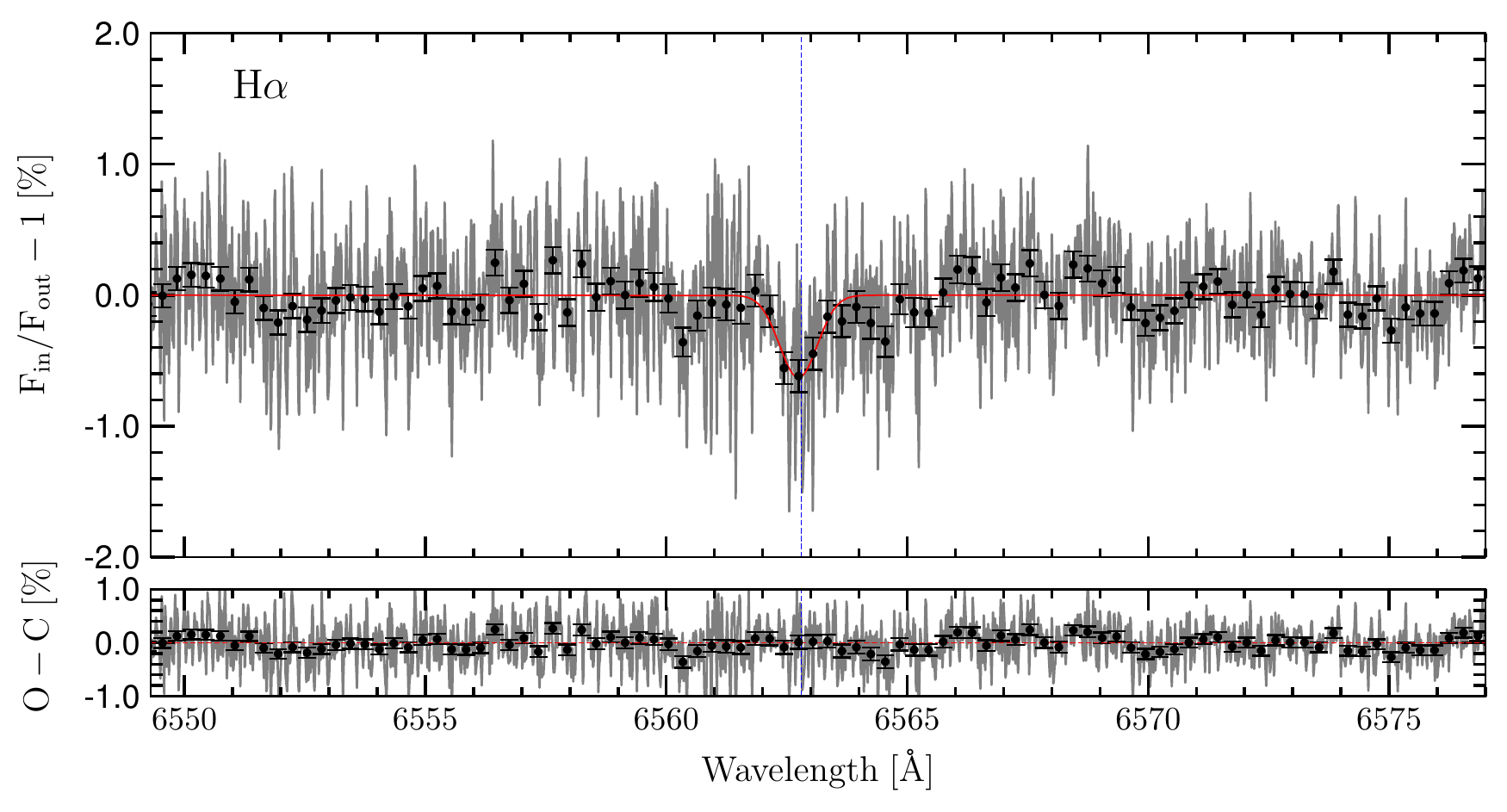}
\caption{Transmission spectrum of MASCARA-2b atmosphere in the region of H$\alpha$. Top panel: atmospheric transmission spectrum (light grey) and binned transmission spectrum by 30 pixels (black dots). The Gaussian fit is shown in red. Bottom panel: residuals of the Gaussian fit. The blue vertical line indicates the expected wavelength position in the planetary reference frame.}
\label{fig:TS_Ha}
\end{figure*}

\begin{figure*}[h]
\centering
\includegraphics[width=1.\textwidth]{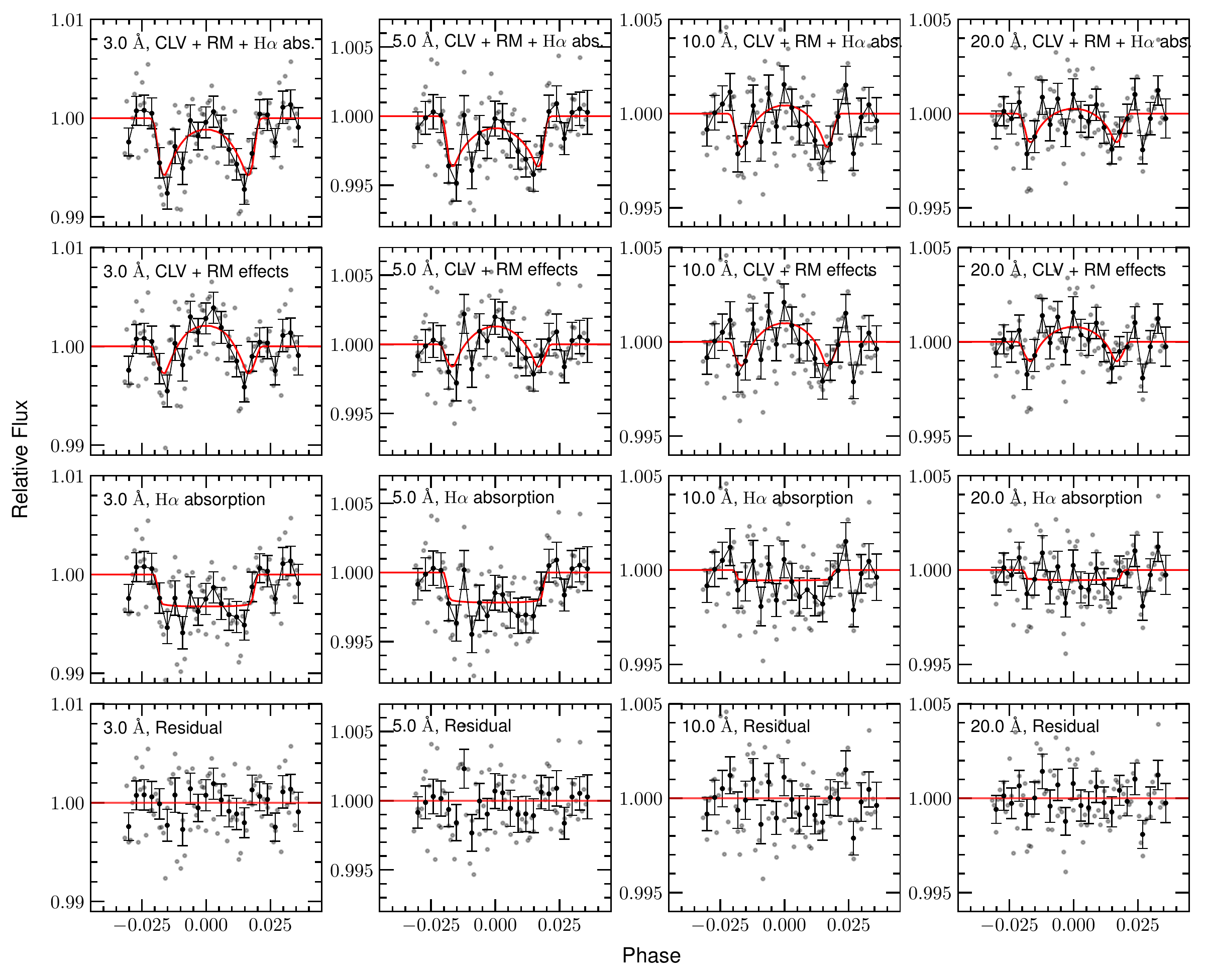}
\caption{Same as Fig.~\ref{fig:TLC_NaI} but for the H$\alpha$ line. In this case, larger bandwidths are used: $3.0~\mathrm{\AA}$ (first column), $5.0~\mathrm{\AA}$ (second column), $10.0~\mathrm{\AA}$ (third column), $20.0~\mathrm{\AA}$ (fourth column).}
\label{fig:TLC_Ha}
\end{figure*}

Following the same methods, we analyze the region around the H$\alpha$ line ($\lambda6562.80~\mathrm{\AA}$) in the MASCARA-2b spectrum (see the 2D representation of the spectra in Appendix~\ref{appendix_2D}). In the telluric correction, we subtract the sky spectrum to correct for possible telluric H$\alpha$ emission. The H$\alpha$ line observed in MASCARA-2 spectra is more than $50~\mathrm{\AA}$ ($5000$ pixels) broad, for this reason the transmission spectrum is binned by $30$ pixels in this case (see Fig.~\ref{fig:TS_Ha}) and larger bandwidths ($3.0~\mathrm{\AA}$, $5.0~\mathrm{\AA}$, $10.0~\mathrm{\AA}$, and $20.0~\mathrm{\AA}$) are used to calculate the transmission light curves (see Fig.~\ref{fig:TLC_Ha}), with $12.0~\mathrm{\AA}$ reference passbands at $\mathrm{B}=[6480.00-6492.00]~\mathrm{\AA}$ and $\mathrm{R}=[6628.00-6640.00]~\mathrm{\AA}$.

A Gaussian feature with a contrast of $0.63\pm0.09\%$ and a FWHM of $0.95\pm0.16~\mathrm{\AA}$ is observed in the H$\alpha$ position ($6562.74\pm0.08~\mathrm{\AA}$), and the measured absorption depth is $0.378\pm0.056\%$ for a $1.5~\mathrm{\AA}$ passband (more values are presented in Table~\ref{tab:AD_valuesHa}). As for the Na I, the H$\alpha$ transmission light curves present a strong RM shape for small passbands which slowly disappears for larger bandwidths (see Fig.~\ref{fig:TLC_Ha}). This RM effect is compensated at about $\sim{50.0}~\mathrm{\AA}$ central passbands, when the stellar line is totally encompassed. However, in this case the possible H$\alpha$ absorption is attenuated in the large passband and the transit light curve becomes flat. The H$\alpha$ absorption depths measured in the transmission light curves measured with different passbands, after correcting for the RM+CLV effect, are presented in Table~\ref{tab:AD_valuesHa}. Similar to the case of Na I absorption, these values are in agreement with the measured absorption depth values from the final transmission spectrum, giving strong confidence in our results.

\renewcommand{\thefootnote}{\fnsymbol{footnote}}
 \begin{table}[h]
\centering
\caption{Summary of the measured relative absorption depth in [\%] in the H$\alpha$ line of the final transmission spectrum and the transmission light curves (TLC) of MASCARA-2b for different bandwidths.}
\begin{tabular}{lcc}
\hline\hline
\\[-1em]
Bandwidth &H$\alpha$& TLC$^($\footnotemark[3]$^)$ \\ \hline
\\[-1em]
$0.188~\mathrm{\AA}$ & $0.521\pm0.159$ &$-$\\ \hline
\\[-1em]
$0.375~\mathrm{\AA}$&$0.678\pm0.110$ &$-$ \\ \hline
\\[-1em]
$0.75~\mathrm{\AA}$ & $0.594\pm0.078$  &$-$ \\ \hline
\\[-1em]
$1.5~\mathrm{\AA}$& $0.378\pm0.056$ &$-$\\ \hline
\\[-1em]
$3.0~\mathrm{\AA}$ &  $0.235\pm0.040$ & $0.311\pm0.044$ \\ \hline
\\[-1em]
$5.0~\mathrm{\AA}$&  $0.206\pm0.031$&$0.213\pm0.034$\\ \hline
$10.0~\mathrm{\AA}$&  $-$&$0.054\pm0.023$\\ \hline
$20.0~\mathrm{\AA}$&  $-$&$0.052\pm0.023$\\ 
\hline\hline
\end{tabular}\\
\begin{tablenotes}
\item Notes. $^($\footnotemark[3]$^)$ Absorption measured in the final transmission light curves (see third row of Fig.~\ref{fig:TLC_Ha}).
\end{tablenotes}
\label{tab:AD_valuesHa}
\end{table}
\renewcommand{\thefootnote}{\arabic{footnote}}

The same control distributions to those presented for Na~$\mathrm{I}$ are performed for H$\alpha$. For a $0.75~\mathrm{\AA}$ bandwidth we measure absorption depths of $0.016\pm0.141\%$, $0.075\pm0.110\%$, $0.136\pm0.085\%$ and $0.387\pm0.078\%$, for the "out-out", "in-in", "even-odd" and "no planet RV" samples, respectively. In this case, when the planet RV correction is not applied we measure a significant absorption depth. This is probably because H$\alpha$ is a very broad line and the planetary signals partially overlap even if the planet RV is not corrected.

%-----------------------------------------------------------------

%

\subsection{(Non-)detections of H$\beta$, H$\gamma$ and Mg $I$ ?}
\label{sec:otherlines}

Following H$\alpha$ we analyzed the H$\beta$ ($\lambda4861.28~\mathrm{\AA}$), H$\gamma$ ($\lambda4340.46~\mathrm{\AA}$) and Mg~$\mathrm{I}$ ($\lambda4571.10~\mathrm{\AA}$) regions, with no clear signals observed in any case (see Fig.~\ref{fig:Aotherlines1} in the Appendix~\ref{appendix_lines}). 

We measure the absorption depth in the expected line position using a central bandwidth of $0.75~\mathrm{\AA}$. For H$\beta$ we use reference passbands of $\mathrm{B}=[4782.0, 4794.0]$ and $\mathrm{R}=[4926.0, 4938.0]$, measuring an absorption depth of $0.079\pm0.059~\%$. Using reference bandwidths of $\mathrm{B} = [4282.00, 4294.00]$ and $\mathrm{R}=[4366.0, 4378.0]$ for the H$\gamma$ line we get an absorption depth of $0.099\pm0.077~\%$. Finally, for Mg I, with $\mathrm{B}=[4492.00, 4504.00]$ and $\mathrm{R}=[4636.00, 4648.00]$, the measured absorption depth is $0.002\pm0.034~\%$. 

For H$\beta$ and H$\gamma$ the transmission spectrum is not totally flat but a lot of noise is concentrated in the expected line positions. Increasing the S/N by co-adding more transit observations is needed to determine the presence of these planetary absorption features.

\section{Modeled Na~$\mathrm{I}$ and H$\alpha$ temperature profiles}

Using high resolution spectroscopy we are not only able to detect chemical species in the atmosphere of exoplanets but also to resolve the spectral lines. When this happens, if the S/N of the final transmission spectrum is high enough, we can adjust isothermal models to different parts of these lines, whose origins reside in different layers of the exoplanet's atmosphere, and reconstruct the temperature profile.

To measure the temperature versus altitude profile from the Na~$\mathrm{I}$ D transmission spectrum we use the atmospheric altitude, $z(\lambda)$, equation given in \citet{2008LecavA}:

\begin{ceqn}
\begin{align}
z(\lambda) = \frac{kT}{\mu g}\ln{\left(\cfrac{\xi\sigma(\lambda)P_o}{\tau_{eq}}\right)\left(\frac{2\pi R_P}{kT\mu g}\right)^{1/2}} + z_0,
\end{align}
\end{ceqn}

where $k$ is Boltzmann's constant, $T$ is the temperature, $\mu$ is the mean molecular weight of the atmospheric composition, $g$ is the surface gravity, $\xi = \xi_i/\xi_H$ is the elemental abundance of the chemical specie $i$ relative to the Hydrogen abundance, $\sigma(\lambda)$ is the absorption cross-section, $P_o$ is the pressure at the reference altitude, $\tau_{eq}$ is the optical depth at the transit radius and $R_P$ is the radius of the planet. The $z_0$ term is independent on the wavelength and can be determined by the altitude in the continuum wavelengths. This relation is derived from a plane-parallel atmosphere, assuming hydrostatic equilibrium and the ideal gas law. 

The cross-section, $\sigma(\lambda)$, of each line is determined modeling the lines as a Voigt profile

\begin{ceqn}
\begin{align}
\sigma(\lambda) = \frac{\pi e^2}{m_e c}\frac{f}{\Delta\nu_D \sqrt{\pi}}H,
\end{align}
\end{ceqn}

where $e$ is the electronic charge, $m_e$ is the electron mass, $c$ is the speed of light, $f$ is the absorption oscillator strength, $H$ is the Voigt profile, which includes thermal and natural broadening, and $\Delta\nu_D$ is the Doppler width, given by 

\begin{ceqn}
\begin{align}
\Delta\nu_D = \frac{\nu_o}{c}\sqrt{2kT/\mu_i},
\label{eq:nuD}
\end{align}
\end{ceqn}

with $\nu_o$ the central frequency and $\mu_i$ the mean molecular weight of the chemical specie being computed. We note that for the Na~$\mathrm{I}$ doublet lines, $\sigma(\lambda)$ is calculated for each line separately and the combined profile is calculated using $\sigma(\lambda) = \sigma(\lambda)_{D2}+\sigma(\lambda)_{D1}$.

For MASCARA-2b, given the presence of H$\alpha$, we assume a mean molecular weight of $\mu=1$. We note that this assumption implies a totally dissociated hydrogen atmosphere, while in truth it could be only partially dissociated. In that latter case, the $\mu$ value would increase and the profile contrast would decrease for similar temperatures. On the other hand, we assume $\tau_{eq} = 0.56$, as it is shown to be mainly constant for planets with $R_P/H{\sim30-3000}$ by \citet{2008LecavA}. We fix $P_o$ to $1~\mathrm{mbar}$. For the Na~$\mathrm{I}$, $\xi$ is fixed to $1.995$x$10^{-6}$ as presented in \citet{huitson2012} and, for H$\alpha$, we estimate $\xi$ with the temperature-dependent relation presented in \citet{2017Huang},

\begin{ceqn}
\begin{align}
\xi \equiv \frac{n_{2p}}{n_{1s}} \simeq 3\frac{n_{2s}}{n_{1s}} = \frac{J_{12}B_{12}}{A_{12}} \simeq 10^{-8}\frac{J_{12}}{10^{-9}\mathrm{erg~cm^{-2}~s^{-1}~Hz^{-1}}},
\label{eq:xi_Ha}
\end{align}
\end{ceqn}

where the peak Ly$\alpha$ intensity is $J_{12} \simeq 0.1 F_{\mathrm{LyC}}/\Delta\nu_D$, and $F_{\mathrm{LyC}} \simeq 9.1$x$10^4~\mathrm{erg~cm^{-2}~s^{-1}}$ \citep{2017Lund} is the flux received by the planet. As defined in Equation~\ref{eq:nuD}, $\Delta\nu_D$ depends on the local temperature.
The values of the Na~$\mathrm{I}$ and H$\alpha$ parameters used here are summarized in Table~\ref{tab:Tprof_param}.

\renewcommand{\thefootnote}{\fnsymbol{footnote}}
\begin{table}[h]
\centering
\caption{Summary of the Na~$\mathrm{I}$ and H$\alpha$ values assumed when fitting $z(\lambda)$.}
\begin{tabular}{lclc}
\hline\hline
                           & \multicolumn{2}{c}{Na~$\mathrm{I}$ doublet}            & H$\alpha$ \\ \cline{2-3} 
                           \\[-1em]
                           & $\mathrm{D_2}$ & \multicolumn{1}{c}{$\mathrm{D_1}$} &           \\ \hline
                           \\[-1em]
$\lambda_o~\mathrm{[\AA]}$ & $5889.951$     & $5895.924$                         & $6562.80$ \\ \hline 
\\[-1em]
$f$                        & $0.6405$ $^($\footnotemark[1]$^)$       & $0.3199$ $^($\footnotemark[1]$^)$                           &    $0.64108$ $^($\footnotemark[2]$^)$       \\ \hline
\\[-1em]
$\xi$                      & \multicolumn{2}{c}{$1.995$x$10^{-6}$ $^($\footnotemark[3]$^)$}                      &   $6.784$x$10^{-6}$ $^($\footnotemark[5]$^)$        \\ \hline
\\[-1em]
$\mu_i ~\mathrm{[AMU]}$      & \multicolumn{2}{c}{$23$}         &  $1$         \\ \hline\hline
\\[-1em]
\end{tabular} \\
\begin{tablenotes}
\item Notes.$^($\footnotemark[1]$^)$\citet{Steck2000}. $^($\footnotemark[3]$^)$ A solar abundance \citep{2003Lodders}. $^($\footnotemark[5]$^)$ Estimated during the temperature fitting with the temperature-dependent Equation~\ref{eq:xi_Ha} \citep{2017Huang}.  $^($\footnotemark[2]$^)$ \citet{problines}.
\end{tablenotes}
\label{tab:Tprof_param}
\end{table}
\renewcommand{\thefootnote}{\arabic{footnote}}

\begin{figure*}[h]
\centering
\includegraphics[width=1.0\textwidth]{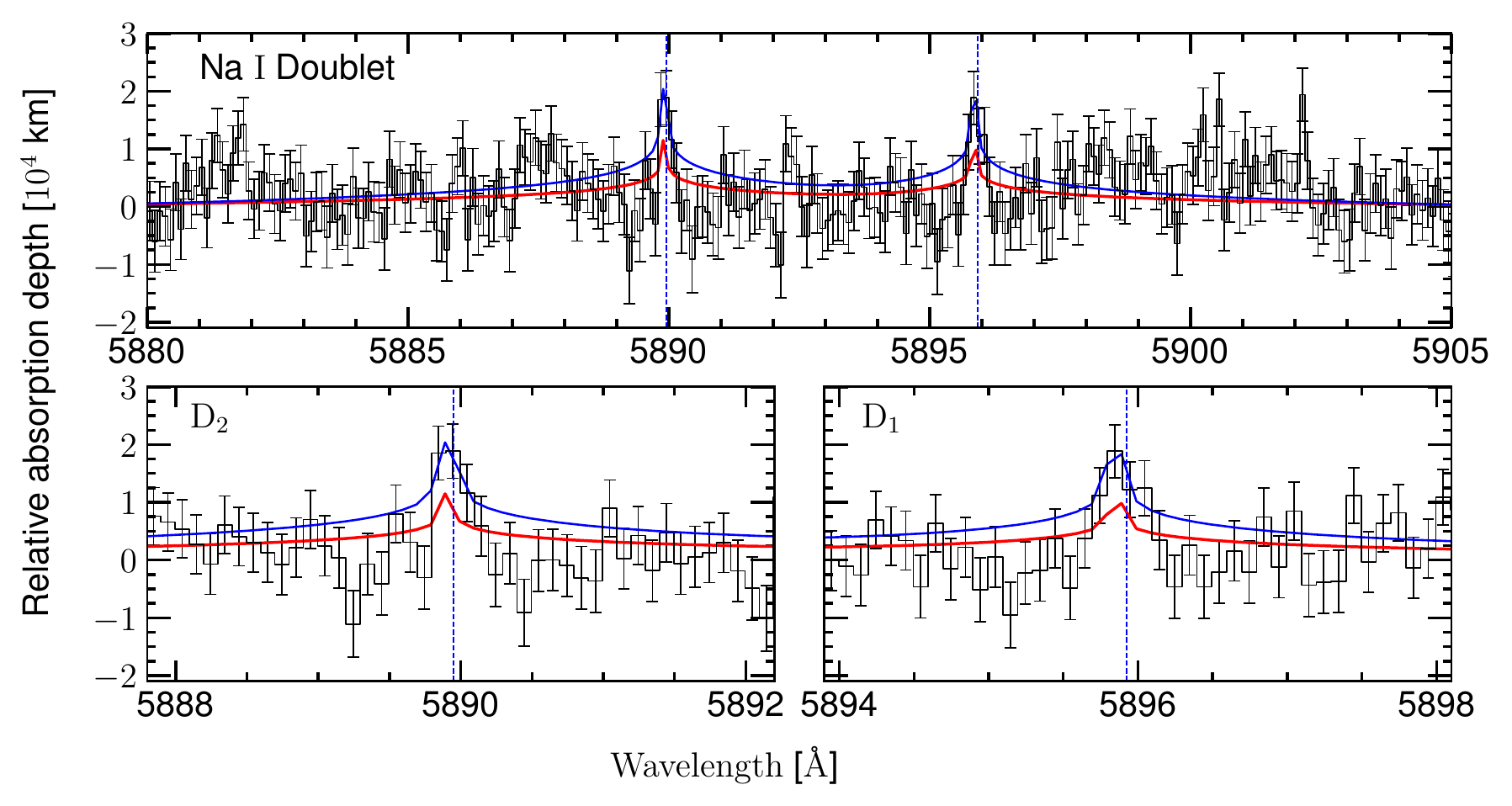}
\caption{Fit of isothermal models to the transmission spectrum of MASCARA-2b in the Na~$\mathrm{I}$ region. The vertical scale is atmospheric altitude in [$\mathrm{km}$] assuming a planet-to-star radius ratio of $R_P^2/R_{\star}^2 = 0.013$. In red we show a model at the equilibrium temperature ($T_{eq} = 2260~\mathrm{K}$), which is adjusted to the continuum. One other model is adjusted to the lines core of each line using the data encompassed for a $0.3~\mathrm{\AA}$ bandwidth centered on the lines peak (shown in blue). For the $D_{2}$ line the best-fit model is at $T=4240\pm200~\mathrm{K}$ and for the $D_1$ line is at $T=4180\pm310~\mathrm{K}$, i.e the same temperature taking into account the uncertainties.  In the top panel we show both Na I lines and the continuum surrounding. In the bottom panels we show each $D_2$ (left) and $D_1$ (right) lines zoomed. Data and models are presented binned by 10 pixels.}
\label{fig:PT_Na}
\end{figure*}

\begin{figure}[h]
\centering
\includegraphics[width=0.50\textwidth]{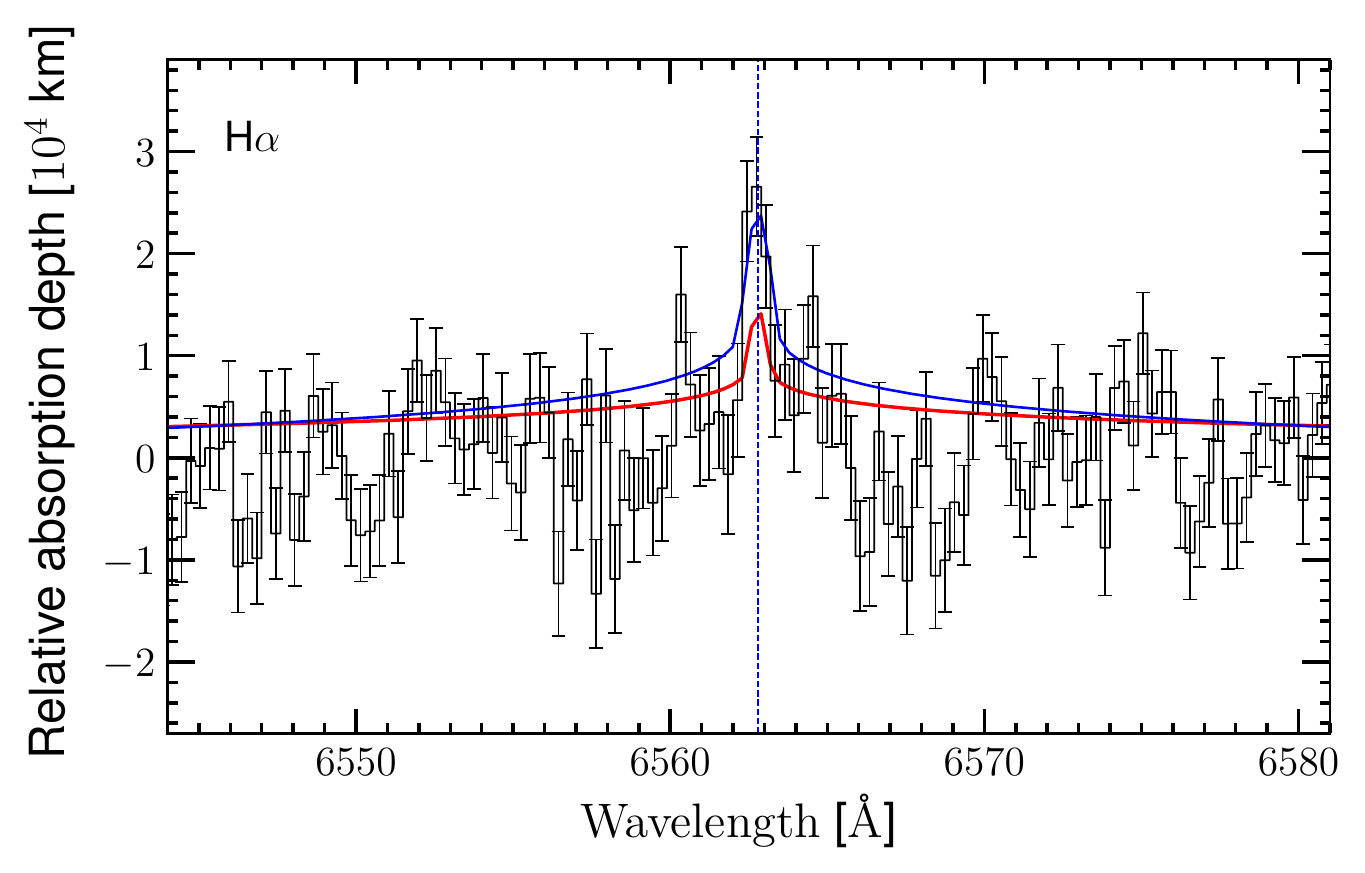}
\caption{Same as Fig.~\ref{fig:PT_Na} but in the H~$\alpha$ region. The isothermal model at the equilibrium temperature ($T_{eq} = 2260~\mathrm{K}$), which is adjusted to the continuum, is shown in red. The blue line represents the model at $4330\pm520~\mathrm{K}$, which is adjusted to the data contained in a $1.2~\mathrm{\AA}$ bandwidth centered to the peak position. Data and models are presented binned by 30 pixels.}
\label{fig:PT_Ha}
\end{figure}

With this information and using the Levenberg-Marquardt algorithm through the \textit{leastsq} tool \citep{scipy}, and similarly to \citet{2015A&A...577A..62W}, we first determine the offset between the models and the data, $z_0$, by fixing the temperature to $T_{eq}=2260~\mathrm{K}$. Once the value is fixed, if the S/N were high enough, we could adjust models to different regions of the spectrum. However, the S/N of MASCARA-2b transmission spectrum is not sufficient to differentiate the wings regions (more transit observations are needed). Still, it is possible to fit the lines core, observing that higher temperatures than $T_{eq}$ are required to explain the contrast measured in the Na~$\mathrm{I}$ doublet and H$\alpha$ lines. In order to fit the temperature of the lines core, we use the data within a $0.3~\mathrm{\AA}$ bandwidth in case of both Na~$\mathrm{I}$ lines, and $0.9~\mathrm{\AA}$ for H$\alpha$ centered to the lines peak, which correspond to the measured FWHM of the lines. We note that each Na $\mathrm{I~D_1}$ and $\mathrm{D_2}$ lines are adjusted separately, with best-fit models at $T=4240\pm200~\mathrm{K}$ and $T=4180\pm310~\mathrm{K}$, respectively, which results in consistent temperatures taking into account the uncertainties. For the H$\alpha$ line the best-fit model presents a temperature of $4330\pm520~\mathrm{K}$. The best-fit models can be observed in Figures~\ref{fig:PT_Na} and \ref{fig:PT_Ha}.

As commented, the best-fit models shown here consider a mean molecular weight of $\mu = 1$, which corresponds to a totally dissociated hydrogen atmosphere. For a molecular hydrogen atmosphere ($\mu=2.3$), the best-fit profiles in the lines core correspond to temperatures larger than $9000~\mathrm{K}$ for the Na~$\mathrm{I}$ ($9250\pm480~\mathrm{K}$) and H$\alpha$ ($9790\pm1770~\mathrm{K}$) lines, i.e. larger than the effective temperature of the host star. However, atmospheres are expected to be heated up to $10~000-20~000~\mathrm{K}$, still producing absorption features in transmission. On the other hand, assuming an atomic hydrogen atmosphere ($\mu=1.3$) these temperatures decrease to $5400\pm200~\mathrm{K}$ and $5530\pm770~\mathrm{K}$ for the Na~$\mathrm{I}$ and H$\alpha$ lines core, respectively, and to ${\sim4000}~\mathrm{K}$ when a totally dissociated hydrogen atmosphere ($\mu=1$) is considered. The MASCARA-2b atmosphere could be only partially dissociated and, in that case, the $\mu$ value would be larger than $1$. However, since the surface gravity of this planet has not been determined yet, and only an upper limit has been estimated, here we assume $\mu=1$, given that if the real $g$ value were lower, it would lead to a similar effect in the temperature as decreasing $\mu$ i.e. increasing the profiles contrast for similar temperatures. Thus, the $9250\pm480~\mathrm{K}$ and $9790\pm1770~\mathrm{K}$ for the Na~$\mathrm{I}$ and H$\alpha$ lines, respectively, obtained assuming $\mu = 1$ and the $g$ upper-limit value, can be considered as upper-limit estimations of the temperatures in these lines.

MASCARA-2b is part of the small sample of planets transiting an A-type star known to date, with effective temperatures higher than ${\sim7000}~\mathrm{K}$. These planets typically receive a large amount of extreme ultraviolet radiation from its host star (${\sim9.1}$x$10^4~\mathrm{ergs~s^{-1}~cm^{-2}}$ in case of MASCARA-2b) , which produce the expansion of their atmosphere and excite the atomic hydrogen to produce H$\alpha$ absorption, and possibly abrasion of the atmosphere \citep{2016Burrier}. H$\alpha$ is commonly used as a stellar activity indicator, being difficult to detect in exoplanet atmospheres, but not for planets orbiting A-type stars, which are not usually active. The measurement of a $0.63\pm0.09~\%$ of H$\alpha$ absorption in the atmosphere of MASCARA-2b corresponds to an effective radius, $R_{\lambda}/R_P$, of $1.20\pm0.04$ and a temperature of $4330\pm520~\mathrm{K}$. These values are obtained with only one transit observation, and considering the strong residuals observed in the H$\beta$ and H$\gamma$ regions of MASCARA-2b transmission spectrum, by co-adding more transit observations we could be able to increase the S/N and study the temperature profile of the planet by observing these other Balmer lines and possibly calculate the lifetime of the atmosphere from escape rate measurements.

%-----------------------------------------------------------------

\section{Conclusions}

We observed one transit of MASCARA-2b, the hot Jupiter orbiting the fourth brightest star with a transiting planet, using the HARPS-North spectrograph. 
In that dataset, we resolve two spectral features centered on the atomic sodium (Na I) doublet position with an averaged absorption depth of $0.172\pm0.029\%$ for a $0.75~\mathrm{\AA}$ bandwidth, measuring line contrasts of $0.44\pm0.11\%$ (D2) and $0.37\pm0.08\%$ (D1), and FWHM of $0.26\pm0.08~\mathrm{\AA}$ (D2) and $0.33\pm0.08~\mathrm{\AA}$ (D1) from the Gaussian fit. The Na I transmission light curves are also calculated, observing a large RM effect for small passbands, which is clearly diluted when the bandwidths are increased. After correcting for the RM and CLV effects using synthetic spectra, we measure a $0.200\pm0.046\%$ Na I transit absorption for a $0.75~\mathrm{\AA}$ passband, consistent with the absorption depth values measured from the final transmission spectrum. We measure the temperature of the Na I lines core, resulting in $T=4240\pm200~\mathrm{K}$ and $T=4180\pm310~\mathrm{K}$ for the $\mathrm{D_2}$ and $\mathrm{D_1}$ lines, respectively, consistent within uncertainties. The S/N of the final transmission spectrum is not sufficient to adjust temperatures in different regions of the lines wings, however, we clearly observe that the equilibrium temperature ($T_{eq}=2260\pm50~\mathrm{K}$) can not define the contrast observed in the Na I lines.

The same method is applied to H$\alpha$, observing a Gaussian feature centered at $6562.74\pm0.08~\mathrm{\AA}$ with $0.63\pm0.09\%$ contrast and FWHM of $0.92\pm0.16~\mathrm{\AA}$. This absorption is also observed in the final transmission light curves, presenting consistent absorption depths to those measured in the transmission spectrum of H$\alpha$. We measure a temperature of $T=4330\pm520~\mathrm{K}$ in the line core, corresponding to an effective radius $R_{\lambda}/R_P$ of $1.20\pm0.04$. Since MASCARA-2b is one of the most irradiated planets to date, we expect the atomic hydrogen to be excited and produce H$\alpha$ absorption. This extreme UV radiation, as for other similar planets transiting A-type stars, can lead to the expansion and abrasion of their atmosphere.

We stress that the results presented here are obtained with only one transit observation, consequence of the brightness of MASCARA-2 and the favorable physical parameters of MASCARA-2b for transmission spectroscopy. Thus more transits are desired to confirm these results. In particular, we observe residual absorption features in the H$\beta$ and H$\gamma$ regions of MASCARA-2b transmission spectrum, but are not statistically significant. More transits would help to  build up enough S/N to determine the presence of these planetary absorptions.

Currently, a small sample of planets are known that transits A-type stars. The study and comparison of these planets, which present different system properties and are irradiated different amounts of energy from their host stars, will help us to learn about their origin and evolution. 

%Additionally, the study of these high-resolution methods is very important for the near future, when high-resolution spectrographs like ESPRESSO on VLT and HIRES on E-ELT will perform transit observations of many exoplanets, making it possible to study their transmission spectra with a higher S/N, characterize rocky-planets and super-Earths. 

\begin{acknowledgements}

Based on observations made with the Italian Telescopio Nazionale Galileo (TNG) operated on the island of La Palma by the Fundación Galileo Galilei of the INAF (Istituto Nazionale di Astrofisica) at the Spanish Observatorio del Roque de los Muchachos of the Instituto de Astrofisica de Canarias. This work is partly financed by the Spanish Ministry of Economics and Competitiveness through projects ESP2014-57495-C2-1-R and ESP2016-80435-C2-2-R. G.C. also acknowledges the support by the Natural Science Foundation of Jiangsu Province (Grant No. BK20151051) and the National Natural Science Foundation of China (Grant No. 11503088). I. Snellen acknowledges funding from the research programme VICI 639.043.107 funded by the Dutch Organisation for Scientific Research (NWO), and funding from the European Research Council (ERC) under the European Union’s Horizon 2020 research and innovation programme under grant agreement No 694513. J. I. G. H. and R. R. L. acknowledge the Spanish ministry project MINECO AYA2014- 56359-P. J. I. González Hernández also acknowledges financial support from the Spanish Ministry of Economy and Competitiveness (MINECO) under the 2013 Ram\'on y Cajal program MINECO RYC-2013-14875. 

\end{acknowledgements}

%   \bibliography{6Yourfile} % your references Yourfile.bib
\bibliographystyle{bibtex/aa.bst} % style aa.bst
\bibliography{bibtex/aa.bib} % your references Yourfile.bib
%\begin{thebibliography}{}

  %\bibitem[Baker(1966)]{baker} Baker, N. 1966,
      %in Stellar Evolution,
      %ed.\ R. F. Stein,\& A. G. W. Cameron
      %(Plenum, New York) 333

   %\bibitem[Balluch(1988)]{balluch} Balluch, M. 1988,
      %A\&A, 200, 58

   %\bibitem[Cox(1980)]{cox} Cox, J. P. 1980,
      %Theory of Stellar Pulsation
      %(Princeton University Press, Princeton) 165

   %\bibitem[Cox(1969)]{cox69} Cox, A. N.,\& Stewart, J. N. 1969,
      %Academia Nauk, Scientific Information 15, 1

   %\bibitem[Mizuno(1980)]{mizuno} Mizuno H. 1980,
      %Prog. Theor. Phys., 64, 544
   
   %\bibitem[Tscharnuter(1987)]{tscharnuter} Tscharnuter W. M. 1987,
      %A\&A, 188, 55
  
   %\bibitem[Terlevich(1992)]{terlevich} Terlevich, R. 1992, in ASP Conf. Ser. 31, 
      %Relationships between Active Galactic Nuclei and Starburst Galaxies, 
      %ed. A. V. Filippenko, 13

   %\bibitem[Yorke(1980a)]{yorke80a} Yorke, H. W. 1980a,
      %A\&A, 86, 286

   %\bibitem[Zheng(1997)]{zheng} Zheng, W., Davidsen, A. F., Tytler, D. \& Kriss, G. A.
      %1997, preprint
%\end{thebibliography}

%\end{document}
%

%
%-------------------------------------------------------------
%               Appendices have to be placed at the end, after
%                                        \end{thebibliography}
%-------------------------------------------------------------
%\end{thebibliography}
\onecolumn
\begin{appendix} %First appendix

\section{2D representations}
\label{appendix_2D}

\begin{figure*}[h]
\centering
\includegraphics[width=0.67\textwidth]{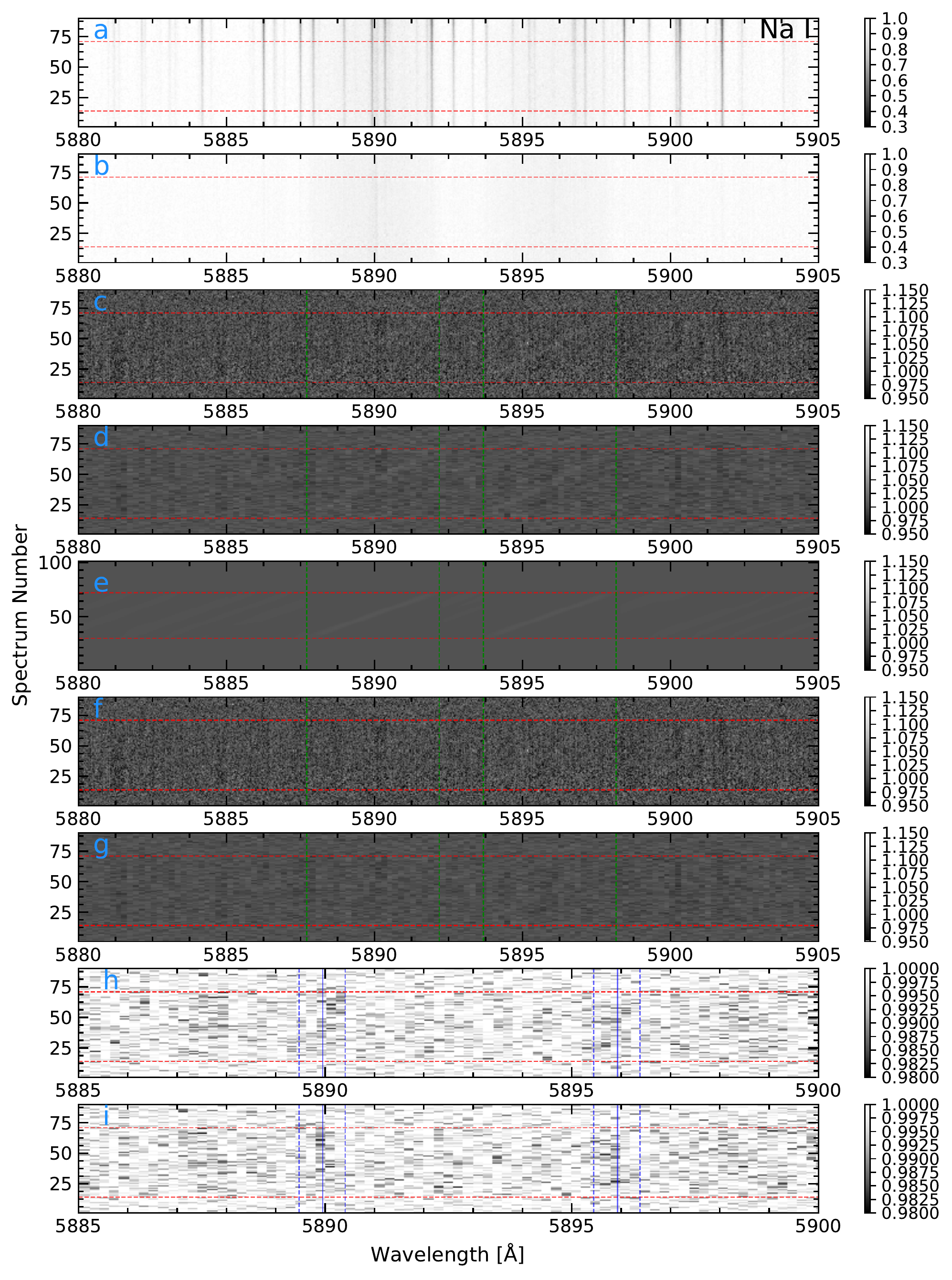}
\caption{2D representation of the MASCARA-2b spectra around Na~$\mathrm{I}$ lines. The vertical axis of each panel represents the sequence number of the observed spectra (the time increases with the spectrum number), being the red horizontal lines the limits between the in- and out-of-transit data. In the horizontal axis we present the wavelength in $\mathrm{\AA}$ and the relative flux values are shown in the color-bar.  \textit{Panel a}: original data after the normalization.  \textit{Panel b}: observed data after the telluric subtraction. \textit{Panels c, d} and \textit{e}: residuals from the in-transit spectra and the Master out spectrum ratio. The RM effect can be clearly observed in these panels. In \textit{ panel d} the data is binned by 20 pixels and in \textit{e} we show the modeled RM residuals (without planetary absorption), obtained by applying to the modeled stellar spectra the same process as data from \textit{panels c} and \textit{d}. \textit{Panels f} and \textit{g}: residuals after correcting each single spectrum presented in \textit{c} for the RM effect (we note in the process presented in Section~\ref{method}, the correction is applied to the final transmission spectrum, after combining all $\mathrm{In/M_{out}}$ with the planet RV correction applied, in order to not affect the planetary absorption). In  \textit{panel g} the data is binned by 20 pixels. The green vertical lines show the limits in wavelength of the RM effect (from $-v\sin i$ to $+v\sin i$). \textit{Panel h}: same as \textit{g} decreasing the wavelength coverage and changing the contrast (see color-bar). \textit{Panels i}: same as the \textit{h} after the planet RV correction. The blue vertical lines of \textit{ panels h} and \textit{i} show the expected  Na~$\mathrm{I}$ position in wavelength due to the maximum and minimum RV of the planet during the transit (dashed line), and the mid-transit value i.e. null RV (solid line).}
\label{fig:Im_Na}
\end{figure*}

\begin{figure*}[h]
\centering
\includegraphics[width=0.67\textwidth]{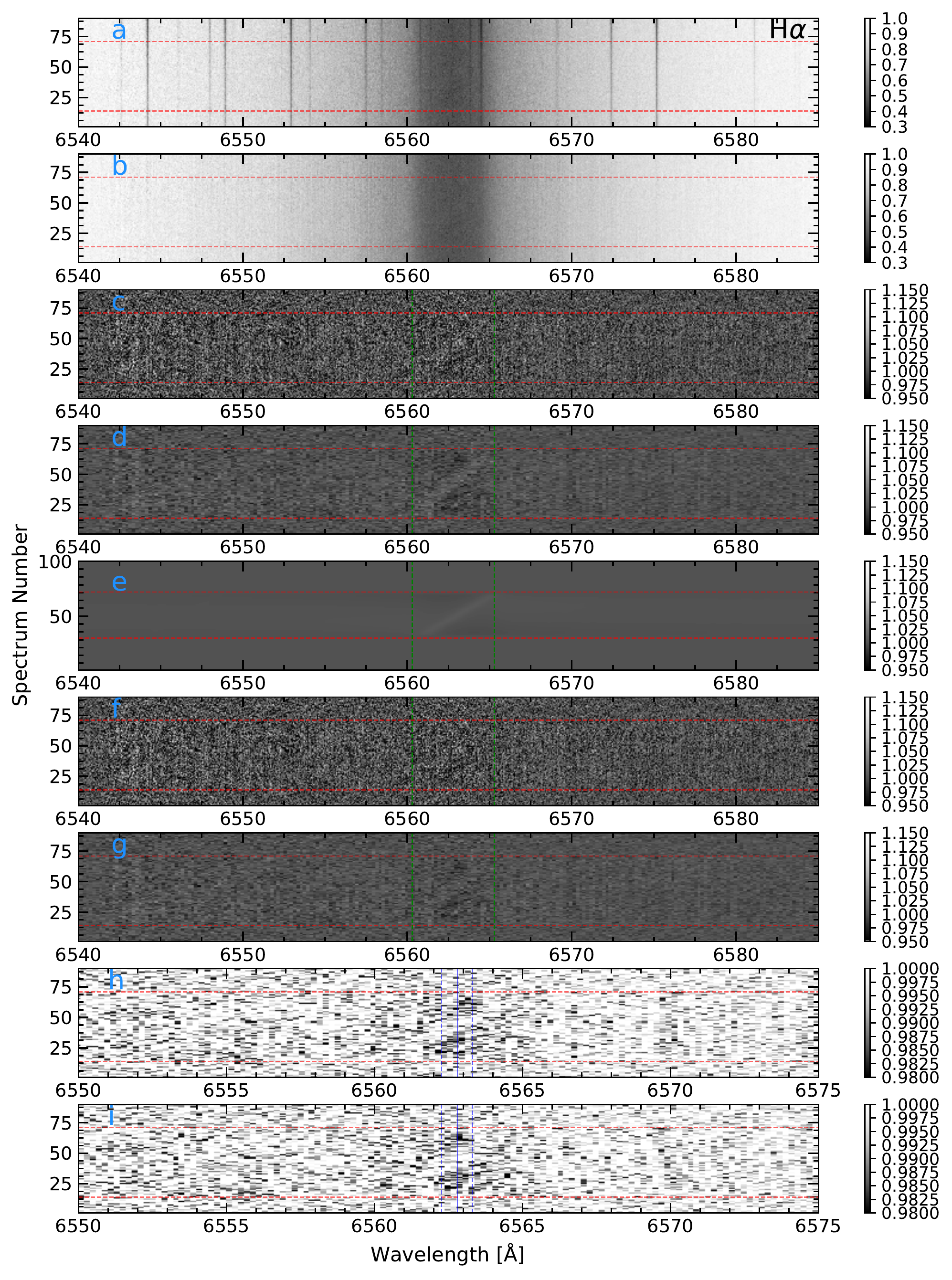}
\caption{Same as Fig.~\ref{fig:Im_Na} but for the H$\alpha$ line.}
\label{fig:Im_Ha}
\end{figure*}

2D representation of the observed spectra and the residuals after some reduction steps presented in Section~\ref{method}, for the Na~$\mathrm{I}$ (Fig.~\ref{fig:Im_Na}) and H$\alpha$ (Fig.~\ref{fig:Im_Ha}) regions. The vertical axis of each matrix represents the sequence number of the spectrum and the horizontal axis is wavelength. A detailed description can be seen in Figure~\ref{fig:Im_Na} caption.

\section{H$\beta$, H$\gamma$ and Mg $\mathrm{I}$ results}
\label{appendix_lines}

As referred in Section~\ref{sec:otherlines}, other regions of the spectra are analyzed by applying the same method presented in Section~\ref{method}. Here we present the resulting transmission spectra around H$\beta$, H$\gamma$ and Mg$~\mathrm{I}$ lines, from which no clear conclusions can be extracted. The S/N needs to be increased by co-adding more transit observations in order to figure out the origin of their residuals.

\begin{figure}[h]
\centering
\includegraphics[width=0.49\textwidth]{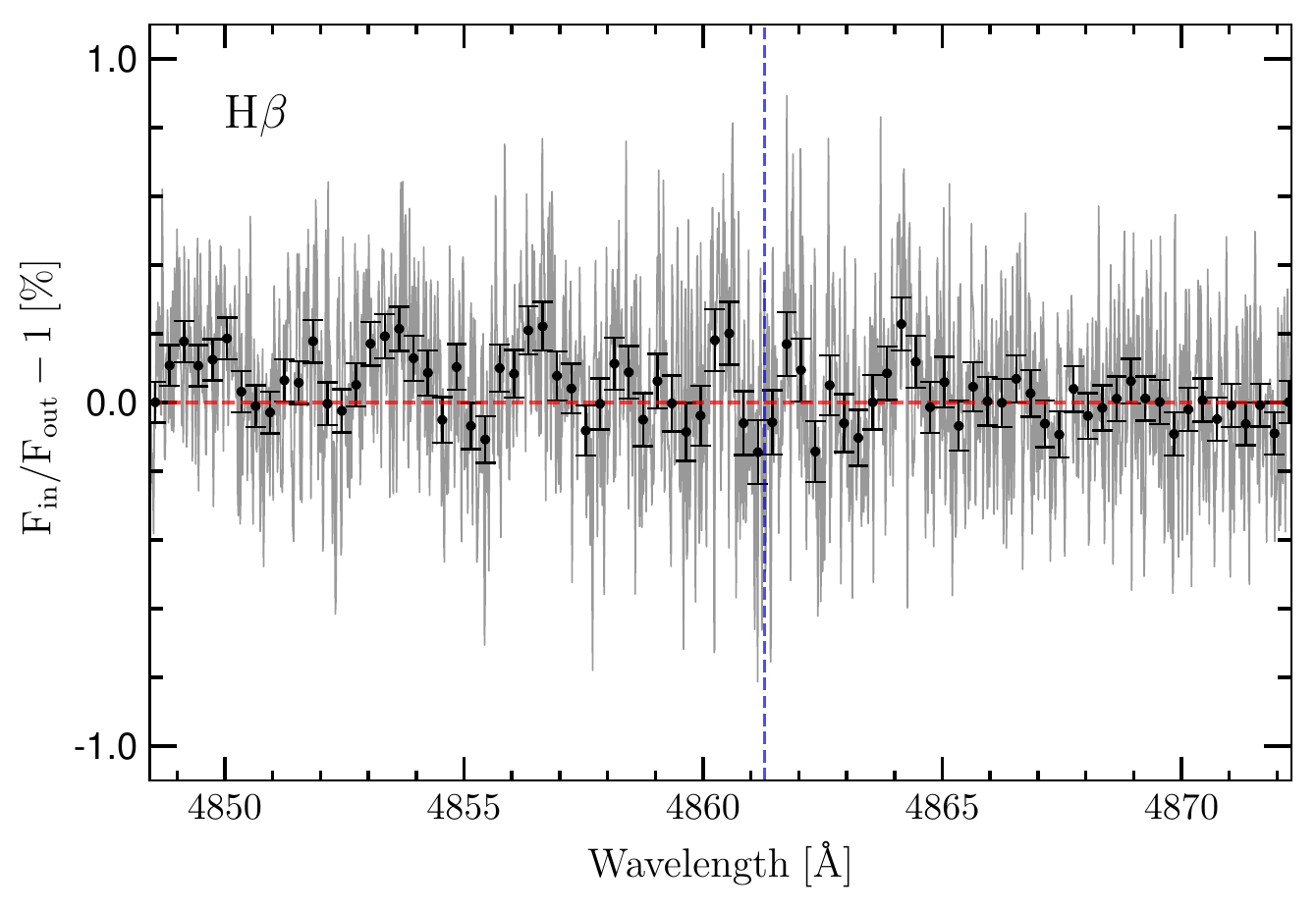}
\includegraphics[width=0.49\textwidth]{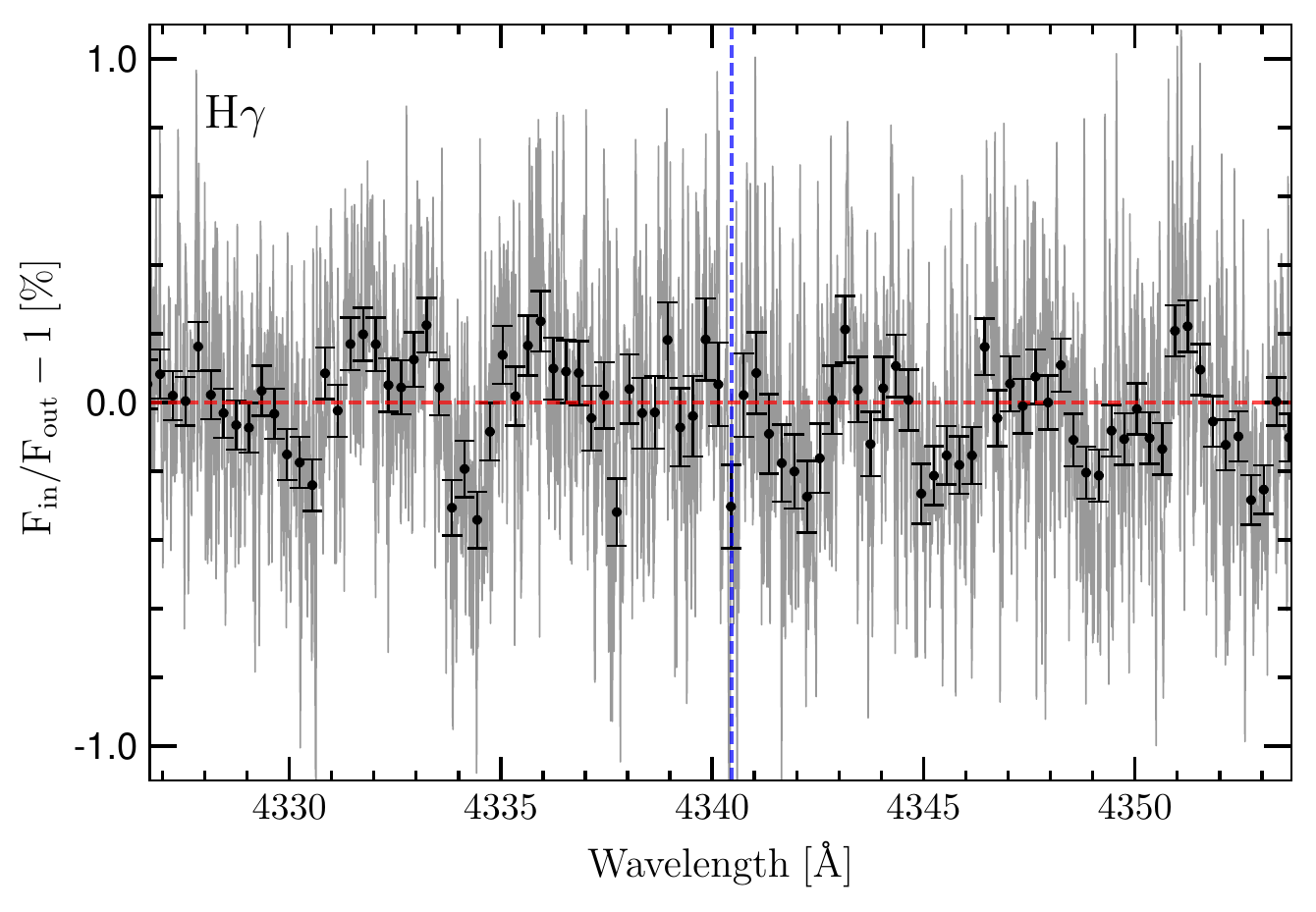}
\includegraphics[width=0.5\textwidth]{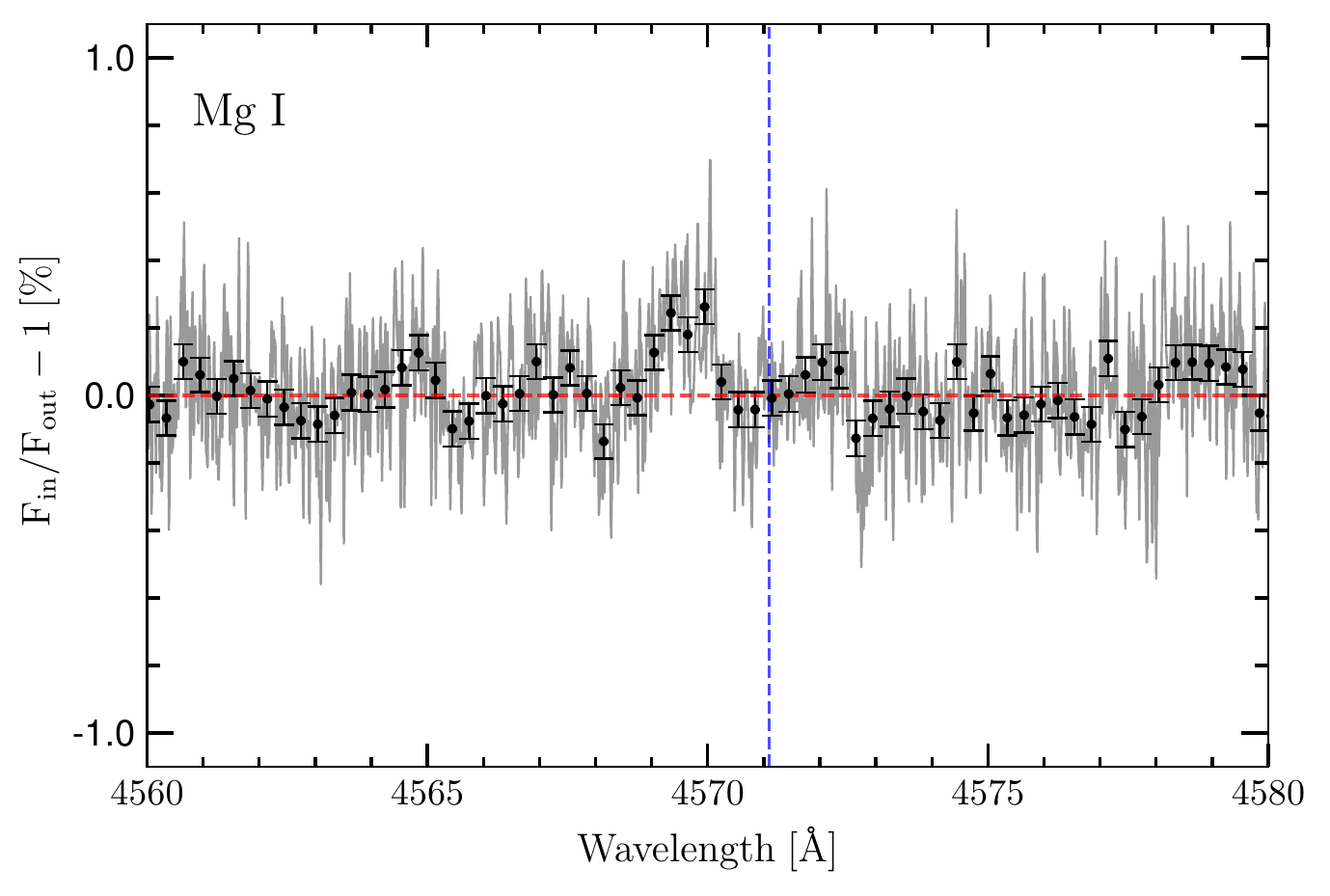}
\caption{Transmission spectrum of MASCARA-2b atmosphere in the region of H$\beta$ (top left), H$\gamma$ (top right), and Mg~$\mathrm{I}$ (bottom). Transmission spectrum (light grey) and binned transmission spectrum by 30 pixels (black dots). The blue vertical line indicates the expected wavelength position in the planetary reference frame.}
\label{fig:Aotherlines1}
\end{figure}

\end{appendix}
\end{document}